\documentclass[review,sort&compress]{elsarticle}

\usepackage{hyperref}

\journal{Modelling and Simulation in Material Science and Engineering}
\usepackage [ngerman, english] {babel}
\usepackage{geometry} 
\usepackage{fullpage} 
\usepackage{fleqn} 
\usepackage{pifont} 
\usepackage{graphicx} 
\usepackage{algorithm}
\usepackage{algpseudocode}
\usepackage{array} 
\usepackage[fleqn]{amsmath} 
\usepackage{amssymb} 
\usepackage{enumerate} 
\usepackage[section]{placeins}
\usepackage{float}
\usepackage[caption = false]{subfig}
\usepackage{tikz}
\usepackage{hyperref} 
\usepackage{cleveref}
\usepackage{caption}
\usepackage{breakurl}
\usepackage{amsmath}
\usepackage{xcolor}
\usepackage[export]{adjustbox}
\usepackage{color,soul}
\usepackage{epstopdf}

\crefname{equation}{Eq.}{Eqs.}
\crefname{figure}{Fig.}{Figs.}
\crefname{algocf}{Alg.}{Algs.}
\crefname{algorithm}{Algorithm}{Algorithms}

\hypersetup {
	plainpages = {false}, 
	breaklinks = {true}, 
	colorlinks = {true}, 
	pdftitle = {A highly efficient 3D level-set grain growth algorithm tailored for ccNUMA architecture}, 
	pdfsubject = {Paper}, 
	pdfauthor = {Christian Mie\ss en}, 
	pdfkeywords = {Grain Growth}, 
	pdfproducer = {pdfLatex}, 
	pdfcreator = {LaTeX (MiKTeX mit TeXnicCenter)} 
  }

\begin{document}

\begin{frontmatter}
\title{A highly efficient 3D level-set grain growth algorithm tailored for ccNUMA architecture}
\author[1]{C.~Mie\ss en} 
\ead{miessen@imm.rwth-aachen.de}
\author[1]{N.~Velinov}
\author[1]{G.~Gottstein} 
\author[1]{L.A.~Barrales-Mora} 

\cortext[cor1]{Corresponding author. Tel: +49 241 80 26899}

\address[1]{Institute of Physical Metallurgy and Metal Physics, RWTH Aachen University, 52056 Aachen, Germany}

\begin{abstract}
A highly efficient simulation model for 2D and 3D grain growth and recrystallization was developed based on the level-set method. The model introduces modern computational concepts to achieve excellent performance on parallel computer architectures. Strong scalability was measured on ccNUMA architectures. To achieve this, the proposed approach considers the application of local level-set functions at the grain level. Ideal and non-ideal grain growth was simulated in 3D with the objective to study the evolution of statistical representative volume elements in polycrystals. In addition, microstructure evolution in an anisotropic magnetic material affected by an external magnetic field was simulated.

\end{abstract}

\end{frontmatter}

\newpage


\section{Introduction}

The evolution of the microstructure during the synthesis, processing and operation of materials has been an important research topic in materials science over decades. Among the different processes that cause modifications to the microstructure, grain growth has recently been put in the focus of interest because of its relevance in nanocrystalline materials. In order to elucidate this physical phenomenon more precisely, researchers have employed statistical analysis on results of computer simulations. Since the first related numerical algorithm was introduced in the 1990s, e.g. \cite{hesselbarth,brakke}, available computational power has grown by orders of magnitude. Despite this fact, microstructure evolution simulation seems to have benefited less from this development. Most algorithms still require prohibitively large amounts of time when simulating representative volume elements (RVE) of microstructure realizations, which are usually required to produce valid statistical results \cite{Kuhbach2016,Kuehbach2014}. Additionally, due to massive computational costs, most of these algorithms are not able to rigorously incorporate the effects of physical phenomena like inclination dependent grain boundary properties, properties of triple lines and quadruple junctions or energy densities.

In order to meet their demand for computational power, researchers have turned to computer clusters. As these resources are expensive to purchase and maintain they are shared between a large number of research groups. This results in limited availability and prioritization of these resources. For these reasons, it is very important to develop algorithms which are efficient and fast, but can still resolve the physical phenomena with comprehensive complexity and high accuracy.

In the specific case of grain growth, several approaches have been introduced in the past. They can be grouped into volume discretized methods, such as phase field models \cite{chengPFM,Stei96,Kazarayanb,Kazarayn2002} or level-set (LS) models \cite{Else09,Else11,Else13,Else14,Esed10,mypaper,Bernacki15,Scholtes2016, Hall13}, explicit methods like vertex models \cite{Kawa89,Weyg98,Barr07,Barr08,Barr10,barrales1,barrales2}, and probabilistic approaches, such as Monte-Carlo-Potts models \cite{Zoel06,Zoel11,Zoel12, SROLOVITZ19861833,SROLOVITZ19882115}. In order to reduce the runtime of these simulations, some of these models were developed to be executed on modern computer architectures \cite{Kuehbach2014, Else13, Scholtes2016, kuehbach15,Yoshihiro07,Yoshihiro08,Nestler2005}. Most algorithms utilize a domain decomposition strategy, which results in explicit communication paths to exchange information across sub-domain walls usually utilizing a message passing interface (MPI). To resolve the original problem without any statistical correction, communication between computational processes is inevitable. The costs depend, however, substantially on the implementation. 

By contrast, the algorithm proposed in \cite{mypaper} divides the microstructure into grains, and duplicates their grain boundaries to create generally independent objects. In this contribution, we present a new implementation of this level-set algorithm for the simulation of grain growth in two and three dimensions. We showed in a previous publication \cite{mypaper} that our model is capable of incorporating a variety of factors influencing grain growth, such as anisotropic GB energies, GB mobilities, and in particular finite junction mobilities in 2D. In this contribution, we discuss the implementation details of the algorithm, and its efficiency for parallel computing applications. An explicit domain decomposition was deliberately avoided and replaced by a dissection along the GBs to facilitate parallelization. As the algorithm implements the grain growth phenomena as competing shrinkage/growth of adjacent grains, data has to be exchanged across GBs to reconstruct the network. The majority of computational effort is localized to individual objects, the grains, to predict their surface evolution. By these simple modifications, it is possible to render the algorithm parallelizable to a great extent as discussed in subsequent sections. 
Our implementation utilizes the OpenMP application programming interface for parallelization. Although OpenMP targets shared memory architectures, we were able to tailor our implementation to a specific ccNUMA architecture. 

More importantly, though, handling grains as individual objects enables us to track the evolution of individual grains and their neighborhood, thus their topological paths \cite{kuehbach15}, which gives the opportunity to study the dependencies of the different mesoscopic features of grain boundaries and their effect on the local topological transitions. The introduced application (GraGLeS\footnote{https://github.com/GraGLeS}) is provided as open source code.

\section{Simulation model}
\subsection{The level-set approach to grain growth}
The level-set (LS) method is a mathematical framework used to depict surfaces and their evolution in time. Since the method was introduced by Osher and Sethian~\cite{osher}, it has been applied to a large number of physical phenomena involving the motion of surfaces in any dimension. This method is capable of coupling the motion of multiple phases separated by sharp interfaces. The enclosing surface of a single region is represented by its own LS function. Thus, an interface is modeled by two congruent surfaces of adjacent regions. In the specific case of grain growth, the LS method implements the physical phenomena as a competing shrinkage of adjacent grains and automatically solves any topological transition and the extinction of entire phases or grains.

\subsubsection{Level-set framework}
\label{framework}

A LS function is a real-valued function defined on the domain $\Omega$. The surrounding GB ($\Gamma$) of an individual grain at time $t$ can be described by means of the LS function $\phi(t,x)$:
\begin{equation}
\Gamma := \{ x  \in \Omega \,| \, \phi(0,x)=0\} \, \subset \, \Omega.
\label{zero level set}
\end{equation}  
This set describes the level zero isosurface of the LS function. Instead of computing the motion of an explicit parametrization of the GB´s, the method tracks the evolution of the LS function $\phi$ and its particular isosurface with level zero (the GB), namely $\Gamma$. 

There are infinitely many possible functions that fulfill the condition in \cref{zero level set}. By convention, if $x$ is inside the grain, $\phi$ will be positive, otherwise negative. To obtain a unique and regular definition of $\phi(t,x)$, another constraint is introduced, namely:
\begin{equation}
\label{gradient}
\left|\nabla \phi \right|=1 , \, \forall x \in \Omega
\end{equation}

An LS function $\phi(t,x)$ that fulfills \cref{gradient} is called a \textit{signed-distance function} (SDF) and will be denoted by $d$ for distinction. This restriction is important to simplify the motion equation of the LS function. In addition, geometric quantities can be directly extracted from the LS function, such as the unit normal vector and the Laplacian curvature, and can be further simplified by means of $d$ (\cref{gradient}):
\begin{subequations}
\begin{equation}
	\label{normal}
	n \, = \, \frac{ \nabla \phi}{\left| \nabla \phi \right|} = \nabla d 
\end{equation}
\begin{equation}
	\label{curvature}
	\kappa \, = \, \nabla \cdot n\, =\, \nabla \cdot \left( \frac{\nabla \phi}{\left| \nabla \phi \right|} \right) = \Delta d
\end{equation}
\end{subequations}

Construction of the total derivative of $\Gamma(t)$ (\cref{zero level set}) leads to the evolution equation for the level-set function \cite{mypaper}:
\begin{equation}
	\frac{\partial \phi}{\partial t} \, + \ v_n \left| \nabla  \phi \right| \, = \, 0
\label{LSmotion}
\end{equation}

Here, $v_n$ denotes the velocity in the normal direction of the isosurface at any point $x \in \Omega$. To illustrate this equation, imagine moving all isosurfaces simultaneously into their normal directions with a velocity according to the local gradient. For our purposes, we assign the motion of an initial surface to the motion of the representing \textit{zero} level-set of the SDF function with time. 

\subsubsection{Modeling grain growth}
\label{graingrowth}

Polycrystalline materials are composed of nearly perfect but differently oriented single crystals. The interfaces separating those crystals or grains are considered as defects within the overall (poly)crystal and are called grain boundaries. Their properties depend on their own atomic structure and have proven themselves to be extremely complex \cite{Brandenburg2014294,Brandenburg2013980,BarralesMora2016179,1757-899X-89-1-012008,barrales-mora2016}. Annealing induces a process of GB migration called grain growth. It reduces the amount of stored energy in the system by elimination of the GBs. This happens when the grain structure of the polycrystal reorganizes and eliminates the defects separating homogeneous regions of atoms. The name is ambiguous, as some grains shrink or completely disappear but the remaining grains grow in a statistical average. This process causes a reduction of the grain boundary area, which is the \emph{defected} area and thus, the source of the free energy.

In a mesoscopic approach to this problem, like the LS method, the complexity of the underlying system is described on a larger length scale than the atomic one \cite{hoffrogge2016kinetic}. On this scale, boundary migration during grain growth can be abstracted as driven by the energy density and therefore, the curvature of the boundary. In turn, the velocity of the migrating boundary can be calculated as:
\begin{equation}
v(t,x(t)) = \mu \, \left(\gamma\, \kappa(x) \, +\, p_B \right)
\label{velocity}
\end{equation}

Here $\mu$, $\gamma(\theta)$, $\kappa$ an $p_B$  denote, respectively, the GB mobility, the GB energy, the local curvature of the GB and any additional driving force such as bulk energy densities resulting from anisotropic magnetic susceptibilities (\cref{magneticE}) or orientation dependent stored elastic energy densities. The entire evolution of a polycrystal can be understood as an energy minimizing process, where \cref{velocity} gives the gradient descent of the 2D energy functional \cite{ZhaoMerriOsher,Zhao_variationalformulation,SROLOVITZ19861833,SROLOVITZ19882115}:
\begin{equation}
E \,=\, \sum_{i<j} \, \textnormal{length}(\Gamma_{i,j}) \, \gamma(\theta)_{i,j} \, +\, \sum_i \, E_i \,\textnormal{area}(\Sigma_i) 
\label{Energyfunctional}
\end{equation}
where $\Sigma_i$ denotes grain $i$.
In the LS approach, a GB $\Gamma_{i,j}$ separating two adjacent grains is interpreted as a part of the surface of each individual phase or grain. Since the global gradient flow can be expressed as a composition of local flows, each grain can be considered as an individual entity \cite{Esed10}. Each entity is represented by one LS function and its isosurface. Owing to this disintegration of the microstructure, voids and overlaps can appear during their independent evolution. A constraint re-composes the microstructure in a procedure generally referred to as \textit{Predictor-Corrector} scheme.

Essentially, this gradient flow describes a certain energy dissipation of a space tessellation. The crucial observation is that the topology of the tessellation changes with time. The topological transitions proceed by the annihilation of GB´s as well as the extinction of complete regions (grains). This leads to certain challenges in the modeling process, e.g. Vertex models need explicit instructions to solve such transformations. By contrast, the LS method has no need for such computational instructions as it automatically handles these transitions with an additional \emph{corrector step} (cf. \cref{predictor} described below). This step balances the predicted motion of free surfaces of individual grains and resolves the equilibrium state at the junctions.

\subsubsection{Algorithm}
\label{subSec:NumericalScheme}

There has been a considerable amount of research in the development of the LS method for grain growth \cite{Else09,Else11,Else13,Else14,Esed10,mypaper,Hall13,Bernacki15,Scholtes2016}. This contribution focuses on an efficient implementation of this method and demonstrates some surprising results regarding the subsequent gain in computational performance. To begin with, the numerical scheme with special emphasis on the implementation demands will be introduced.

\begin{figure}[h]
\begin{minipage}{\textwidth}
\captionsetup[subfloat]{captionskip=1pt} 
	\centering
	\subfloat[]{\label{distancefunction} \includegraphics[width = 0.5\textwidth,frame ]{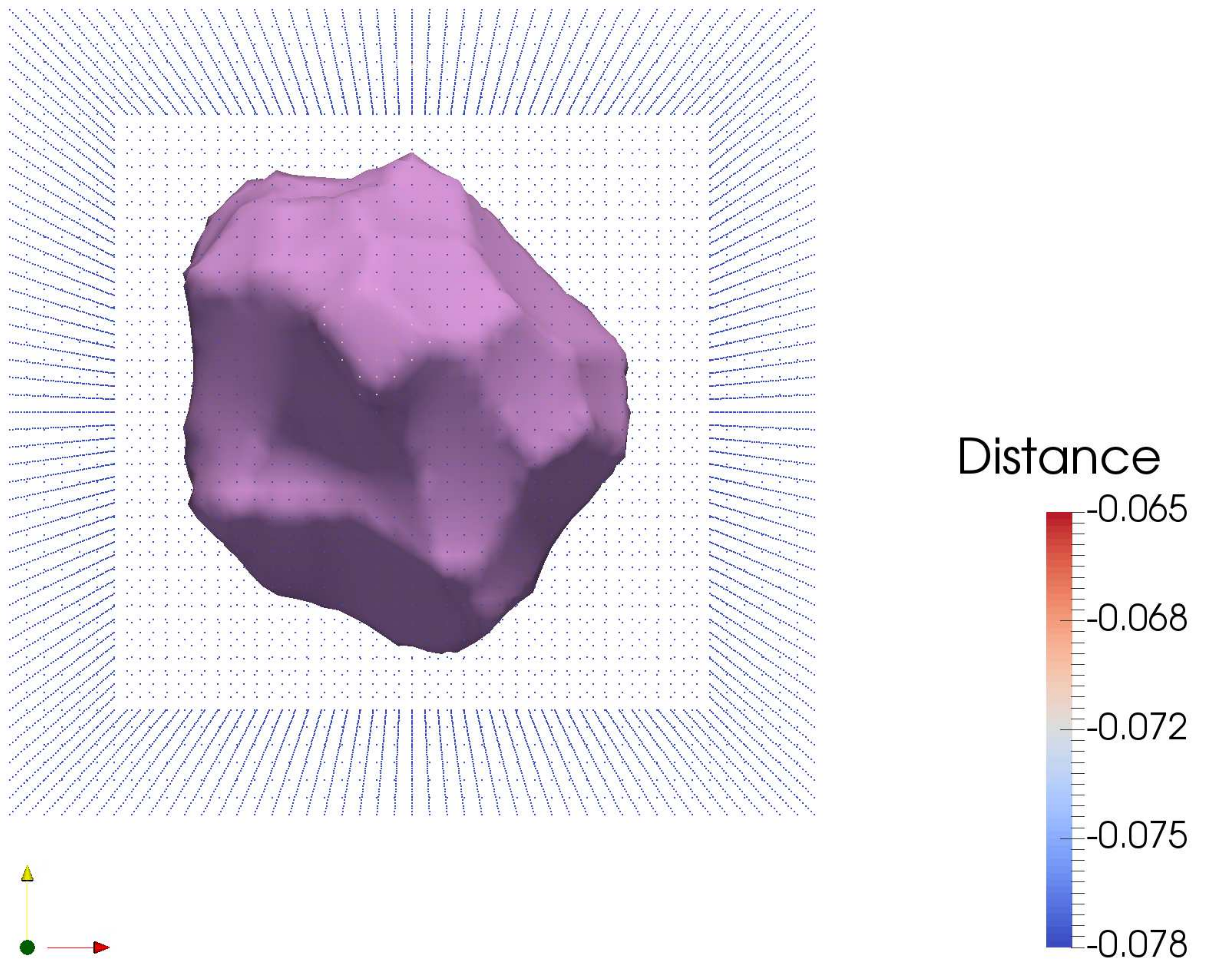} }
	\subfloat[]{\label{mgcontour} \includegraphics[width = 0.5\textwidth, frame]{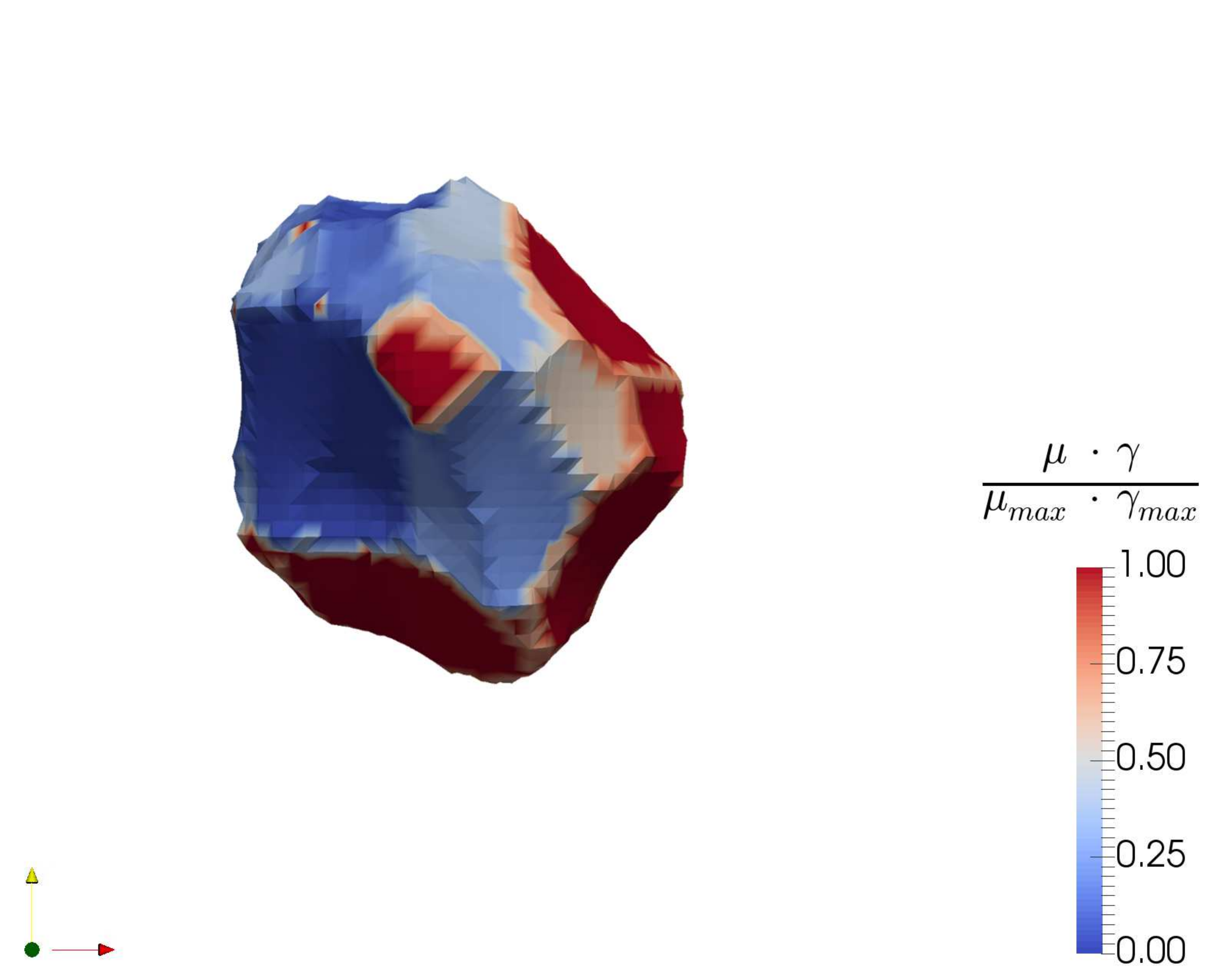} }
	\caption{a) A single 3D grain is described by its own level-set function. The GB is rendered as isosurface with level zero. Here the local domain $\Omega_i$ has a size of $47 \times 47 \times 47$ grid points, whereas there is the need to keep at least 7 additional grid points into each spatial direction. The distance function was only evaluated in a tubular region enclosing the GB ($\left| \phi_i \right| \leq \epsilon$), outside it was restricted to $\epsilon = -7 \cdot h = -0.078$, where $h$ is the grid spacing. The palette indicates the value of the signed-distance function. b) The same isosurface is colored by its physical GB character $\mu\cdot\gamma$. Triple junctions, GBs and even quadruple junctions can be identified by our scheme and hence can be associated with different physical properties.}

\label{variousdsf}
\end{minipage}
\end{figure}

In this scheme, a polycrystal encompasses the region $\Omega := [0,1]^D \subset \mathbb{R^D}$, where $D$ denotes the dimension, in particular 2 or 3. We assume that there is a tessellation $S$ of $\Omega$ that represents a polycrystal. Each sub-region $S_i$ is described initially as a polytope and represents a grain of this polycrystal. For a numerical description, each element $S_i \in S$ owns a LS function $\phi_i(x,t)$ (\cref{zero level set}) on a discrete domain $\Omega_i \subset \mathbb{N}^D$. In order to have a specific amount of points within each grain on average, the discretization of the computational domain $\Omega$ is first defined. Based on the overall dimension of the represented grain, $\Omega_i$ is extended by a certain offset of additional grid points to encompass the entire GB of the corresponding grain by means of its LS function $\phi_i(x,t)$ (\cref{distancefunction}).

The level-set function $\phi_i$ is initialized to fulfill \cref{gradient}, thus the scheme represents each grain by its SDF ($d_i$). The corresponding grid $\Omega_i$ must also be size-adaptive to accommodate the change in the size of the grain.
Having defined $d_i \, \forall i \in I$, the GB migration of a D-dim ($D\in\{2,3\}$) polycrystal can be computed with \cref{predictor} proposed in \cite{Else09,Else11,Else13,mypaper} and shown below.

\begin{algorithm} [h!]
\caption{Level-set algorithm for grain growth}
\label{predictor}
\begin{algorithmic}[1]
\State \textit{Initialization:} Create all $\Omega_i$ for all subset $S_i$ of the tessellation $S$  
\While{$t<=T_{end}$} 
\State \textit{Predictor step:} Compute the motion of individual grains
\State \textit{Corrector step:} Remove voids and overlaps
\State \textit{Re-initialization:} Compute distance functions to new interface positions 
\EndWhile
\end{algorithmic}
\end{algorithm}

\subsubsection{Predictor step}
\label{subsec:predictorstep}
In order to simplify the driving force defined in \cref{velocity}, we first normalize the anisotropic velocity of the GB ($\mu \gamma = 1$) to simplify \cref{velocity} as: 
\begin{equation}
	\label{Heatequation}
	\frac{\partial d}{\partial t} \, - \, \Delta  d  \,=\, 0
\end{equation}

This differential equation describes the heat flow with $d_i$ corresponding to the temperature distribution. The Green's function for this problem is known and given by the \textit{Gaussian kernel} $G(t,x)$ \cite{Evans} :
\begin{equation}
	G(t,x)\, =\, \frac{1}{(4\pi t)^{1/D}} \, e^{-\left| x\right|^2 / 4\pi}
	\label{Gaussian}
\end{equation}

Convolving the distance function with the kernel gives a solution to initial value problem $d(0,\cdot)=g_0 $. In turn the application of the convolution will generate an evolution of the interface:
\begin{equation}
\label{solution}
d(t,x)\, =\, G(t,x) \, \ast \,d(0,x)
\end{equation}
 
Utilizing this approach, motion by mean curvature of an isosurface can be computed in any dimension. Note that this convolution can be efficiently solved in Fourier space. To extend this approach to a more complex driving force, which is weighted by a local anisotropy of the GB ($w$), we developed a correction procedure in \cite{mypaper}. For this, we assumed that the energy and the mobility at any point on the GB is known and that each grid point has adopted this feature from the closest point on the GB. The subsequent correction evaluates two states of the level-set function, an adequate signed-distance function ($d_k$) and a convolved level-set function ($d_k^p$):
\begin{equation}
\label{eq:velocorrect}
\begin{aligned}
	&v_n\,\stackrel{\cref{hoehenvari}}{=} \,-( d_k \, - \, d_k^p )(x)/ \Delta t  := l_t  \\
	&d_k^{p} (x,t)\,=\, d_k (x,t) \,+\,(v_n w \Delta t) + (\mu \Delta E \Delta t)
\end{aligned}
\end{equation}

Finally, we exploit the simple connection between a metric on surfaces and a metric on level-sets which is given by:
\begin{equation}
	l_t \, :=\, \frac{\partial \phi}{\partial t} \, =\, -v_n \left| \nabla \phi \right|
	\label{hoehenvari}
\end{equation}

This corresponds to the distance each point of the GB is moved into its normal direction during the time $\Delta t$. The velocity is corrected by the weights $w$ representing the GB anisotropy \cite{Nemitz}. $w$ already consists of the GB mobility $\mu$ and energy $\gamma$, thus $\mu$ has to be applied separately on the bulk energy (\cref{eq:velocorrect}). The quantities affecting the curvature are normalized to the interval $(0,1]$ using the given maximal physical values for mobility $\mu_{max}$ and energy $\gamma_{max}$ for normalization. In the simplest case, the bulk energy is homogeneously distributed for each grain, whereas the driving force stems from the gradient descent of both potentials across the GB. These energies (magnetic or stored elastic) can also be given in physical units but have to be normalized to account for the dimension of the sample and the physical problem at hand.

To derive the correct normalization of the bulk energy, we consider a spherical grain with radius $r$. $L\left(D\right)$ denotes the dimension of the domain in one spatial direction represented by the unit interval $I=\left[0,1\right]$. We replaced all variables in \cref{velocity} by their correspondent normalized equivalents. Note that an asterisk $^*$ will always indicate that a variable is contained in the normalized space, whereas a normal notation will assign the physical equivalent. This spherical grain is assumed to be in an equilibrium of forces resulting in $v_n=0$ (\cref{velocity}):
\begin{equation}
\kappa \,=\, \frac{2}{r} \, =\,\frac{2}{r^* \cdot L(D)} \,=\, \frac{p_B}{\gamma}
\label{eq:}
\end{equation}

where $r^* \in (0,1)$. Thus, we deduce a normalization coefficient $c_B$ for a driving force $p_B$ originating from a bulk energy given in physical units:
\begin{equation}
c_{B} \,=\, \frac{L(D) }{\gamma_{max}} 
\label{bulk normalization}
\end{equation}

In addition, it is very important to connect the real (physical) time increment $\Delta t$ with the internal time increment $\Delta t^*$. Again, we replace the physical quantities by their corresponding ones in the normalized space to deduce the connection:
\begin{align}
\Delta t \quad &= \quad \frac{\Delta r}{v_n} \quad= \quad \frac{\Delta r^* \cdot L(D)}{v_{n}^*\cdot\frac{1} {L(D)} \cdot \mu_{max}\cdot \gamma_{max}}\\
&= \quad  \frac{v_{n}^* \cdot \Delta t^* \cdot L(D)^2}{v_{n}^* \cdot \mu_{max} \cdot \gamma_{max}}\quad  = \quad \frac{\Delta t^* \cdot L(D)^2}{\mu_{max}\cdot \gamma_{max}}
\end{align}

This relation is very advantageous as it enables comparing simulation results with experimental findings and analytical solutions of certain growth phenomena.

\subsubsection{Corrector step}
\label{comparison step}

Voids and overlaps appear during the predictor procedure because the grains are modeled as independent objects, whose evolution leads naturally to these features. The \emph{corrector step} redistributes the critical grid  points in those regions among all competing grains to reach an equilibrium state. This procedure ensures that the computational domain stays entirely covered by disjunct phases. More importantly though, this step automatically handles any topological transition and the elimination of entire grains. In practice, the comparison step is done by matching the predicted level-set functions $d_k^{p}, \, k\in \left\{1 \dots M \right\}$ of adjacent grains:
\begin{equation}
	d_k^c (x,t+\Delta t) \,= \,  \frac{1}{2} \,(d_k^{p} (x,t+\Delta t) \,- \,max_{j\neq k} \, d_j^{p} (x,t+\Delta t))
\end{equation}

\subsubsection{Re-initialization}
\label{subsec:Re-initialization}
Finally, a re-initialization of all level-set functions $d_k^c$ to the corresponding zero level-sets is needed to maintain numerical stability. Solutions to this problem have been provided in terms of equidistant grids in \cite{Hartmann,Russ00,Sussman,zhao04}. The re-initialization of the signed-distance function faces two problems. First, the position of the interface has to remain invariant. Second, the gradient of the re-initialized level-set function must fulfill $\left| \nabla \phi \right| = 1$ to ensure a correct evaluation of the Laplacian curvature in the subsequent \emph{predictor step} \cref{Gaussian}. The formal complexity of the available methods differs substantially. By investing higher computational costs ($\sim$ 40 times more), higher order accuracy of the resulting signed-distance is possible \cite{Hartmann,Russ00,Sussman}. This re-initialization approach originates from the reformulation of the Eikonal equation as an evolution equation introduced first by Sussman et al. \cite{Sussman}:
\begin{equation}
\phi_{\tau} + S(\phi_{\tau=0}) \left(\left| \nabla  \phi \right| -1 \right)\,=\,0
\label{SussmannODE}
\end{equation}

Here, $\tau$ is an artificial time step size. A major drawback of this approach is that the zero level-set is considerably displaced and the deviation could even increase with the number of iterations used to solve the ODE (\cref{SussmannODE}).
Russo and Smereka \cite{Russ00} improved the stability of the method in such a way that the displacement of the zero level-set is independent of the number of iterations used to solve \cref{SussmannODE}. This was achieved by using only information from one side of the zero level-set in the discretization of the ODE.  

The reformulation by Hartmann et al. \cite{Hartmann} is a modification of the partial differential equation introduced by Sussman et al. \cite{Sussman} and, in particular, improves the second-order accurate modification proposed by Russo and Smereka \cite{Russ00}. The latter authors proposed two formulations of the reinitialization problem, namely, to explicitly minimize the displacement of the zero level set within the reinitialization and to solve the overdetermined problem to preserve the interface locally within the reinitialization. Thus, they proposed a direct update of the distance values next to the zero level-set. Subsequently, the re-initialization starts from the side of the interface providing the most points along this segment of the isosurface.

By contrast, a fast-sweeping algorithm as proposed in \cite{zhao04} reconstructs the corresponding signed-distance values only by first order accuracy but still minimizes the displacement of the zero level set. Starting from a fixed set of distance values next to the interface the distance information is spread into all spatial directions incrementing with the the grid-spacing $\Delta h$. Thus, the resulting distance values are given as \textit{taxicap}\footnote{The \textit{taxicap} metric or the \textit{Manhattan} norm, proposed by Hermann Minkowski in 19th-century Germany, replaces the Euclidean metric, the distance between two points, by the sum of the absolute differences of their Cartesian coordinates. The name traces back to the grid layout of most streets on the island of Manhattan, where the shortest path a car could take between two intersections has a length equal to the intersections' distance in taxicab geometry.}
distances to the closest grid-point next to the interface. The distance calculation of these reference points is exact. 

Such algorithms are usually appraised depending on the accuracy of the reconstructed SDF function related to a well-discretized but complex shape. Alternatively, they are also valued for their computational performance in applications where high accuracy is not required such as in video games. In our application, the shapes are not complex because the grains resemble in most cases polyhedra with curved faces. Here, the challenge are vanishing grains, which are always poorly discretized and thus the evaluation of their curvature is hardly possible. Unfortunately, these grains can influence substantially the growth kinetics of the network as their released volume will be occupied by their adjacent grains and thus, the shrinkage regulates the growth. The effect is somehow inverse to what was found for finite triple junction mobilities, which affect the global kinetics by dragging the shrinkage of the smallest grains close to elimination \cite{barrales1, barrales2, mypaper}.

A comparison between both approaches is given in \cref{redistComp} and is discussed in \cref{sec:numacry}. 
\section{Novel numerical approaches}
\label{sec:numinno}

\subsection{Datastructure}
\label{subsec:datastrc}

In this section, the implementation details which were crucial for the resulting code performance will be presented. They influence both sequential and parallel execution times. The programming language \textit{C++} was selected as it provides user defined memory allocation and allows to structure the program in accordance to the physical model, using the object-oriented paradigm. Additional libraries like \textit{FFTW} \cite{FFTW05}, \textit{Intel MKL} \cite{intelMKL} and \textit{EIGEN} \cite{eigenweb} were used for some mathematical algorithms. 

The input data for the simulation is a tessellation $S$ (as described in~\cref{subSec:NumericalScheme}). For each element $S_i \in S$, a grain object is created. The grain contains physical information (orientation, neighborhood, etc.) and represents the grain in LS space. Hence, the main purpose of the grain object is to keep track of the distance values $B_i := \left\{d_i(x)\, | \,x\in \Omega_i \right\}$ (\cref{distancefunction}). As every step of the algorithm transforms the function $d_i$ in some way, we keep two buffers $B_i^1$ and $ B_i^2$ in each object. One is used as input to the current computation step whereas the other serves as output. Consequently, references to those buffers are switched back and forth after each step of the \cref{predictor}; this is usually called \textit{double buffering}. These buffers have to accommodate size changes of the corresponding grain. To accomplish this, when a buffer grows it is increased with more memory than requested, in order to accommodate future growth. This heuristic, combined with the grid coarsening procedure (cf. \cref{sec:seqopt} below), reduces the relocation of objects in memory substantially.

During the \emph{corrector step}, information has to be exchanged among competing grains to redistribute the space along their boundaries and junctions (\cref{comparison step}). As grains are hosted by surrounding boxes, those boxes give a suitable instrument to compute a preselection of possible neighbors to a certain grain. In the initial time step, an \textit{Rtree} is used to find all overlapping boxes as input for the comparison routine \cite{Guttman:1984:RDI:971697.602266}. Subsequently, the local neighborhood is updated by the evaluation of the points of contact to preselected neighbor candidates. Memory fetches are only performed on the resulting box intersections ($x \in\Omega_i \cap \Omega_j$) with the already identified neighbors. Furthermore, for each point in $\Omega_i$, we locally collect information about the nearest neighboring grain to that point (grain $j  \, |\, \textnormal{max}_{j\neq i} d_j(x) $) in a structure called \textit{ID-field}. For more precise neighbor statistics, the \textit{ID-field} (which is part of the grain object) is evaluated next to the isosurface of level zero i.e. the GB. 

All other subroutines of the algorithm are local (in terms of memory fetches) to the volume of the local grid $\Omega_i$ hence to the grain object itself. As we are aiming at efficient parallel computations, the structure of the grain objects allows very flexible task scheduling since sets of grains can be directly assigned to computational threads.

\subsection{Sequential Optimizations}
\label{sec:seqopt}

In \cref{subsec:predictorstep}, it was mentioned that each point in $\Omega_i$ needs to adopt the physical properties of the nearest point on the GB. Mathematically, this mapping was an easy problem that, nonetheless, required a tremendous numerical effort. The SDF was inappropriate to evaluate this connection as it provided solely scalar information in form of distance values. Therefore, the position of the GB had to be computed first using a marching squares/cubes algorithm depending on the dimension to obtain an explicit representation \cite{marchingCubes}. Hence, the closest snippet of that linear representation to each grid point had to be identified. Unfortunately, there were, to our knowledge, no algorithm in 2D or in 3D that can solve the point to polygon and the point to triangulation problem in sub-linear complexity. In both cases, the complexity is dominated by the cardinality of the set of linear segments or triangles ($O(E)$, where $E$ is the number of edges of the polygon).

In order to reduce the formal complexity, we opted to cluster the linear representations of the grains according to the number of neighboring grains next to each piece of the linearization using the procedure described in \cref{subsec:datastrc}. Utilizing this information about the GB character itself, it was possible to distinguish between segments forming a GB, a triple line, a quadruple junction or even high order junctions. This concept and the utilization of the ID-field lead to a massive reduction of the complexity of the search for the next closest segment on the grain boundary. Additionally, the expensive evaluation was restricted to the set of points enclosed by a narrow tube around the GB. 

As a further sequential optimization of the code, a grid-coarsening routine was applied, which kept the spatial resolution of the grains constant with time. Thus, the grid spacing grew inversely proportional to the population so that the average discretization per grain remained constant. The projection of each SDF to a coarser grid was performed by a tri-linear interpolation scheme. The improvement was substantial as it not only reduced the memory consumption with time but also allowed increasing the time step simultaneously, with proportionality $1/N^2$.

\subsection{Parallelization}
\label{sec:parallel}

\begin{figure}[!h]
\begin{minipage}{\textwidth}
\captionsetup[subfloat]{captionskip=1pt} 
\centering
	\subfloat[]{\label{iterative} \includegraphics[width = 0.5\textwidth]{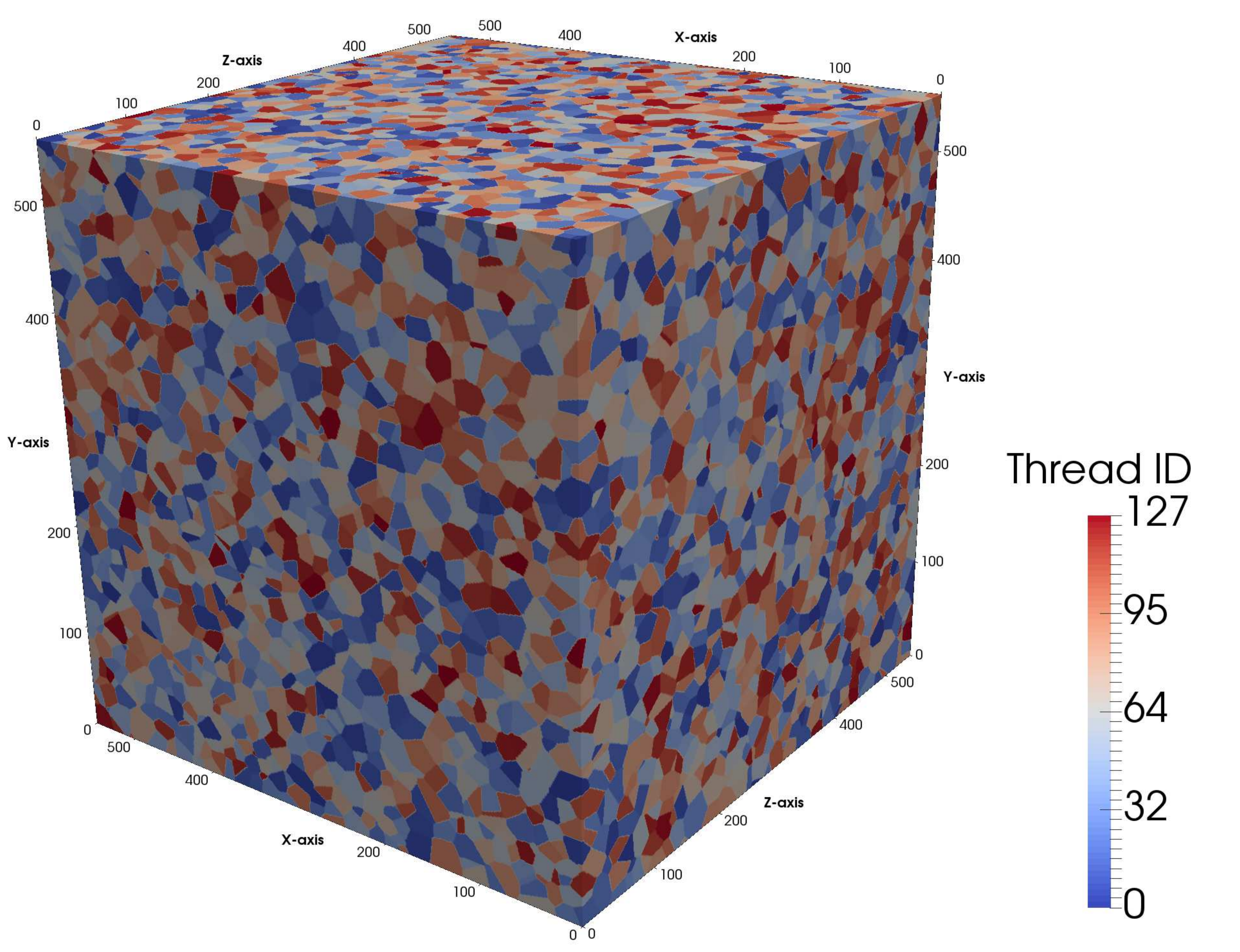} }
	\subfloat[]{\label{squared} \includegraphics[width = 0.5\textwidth]{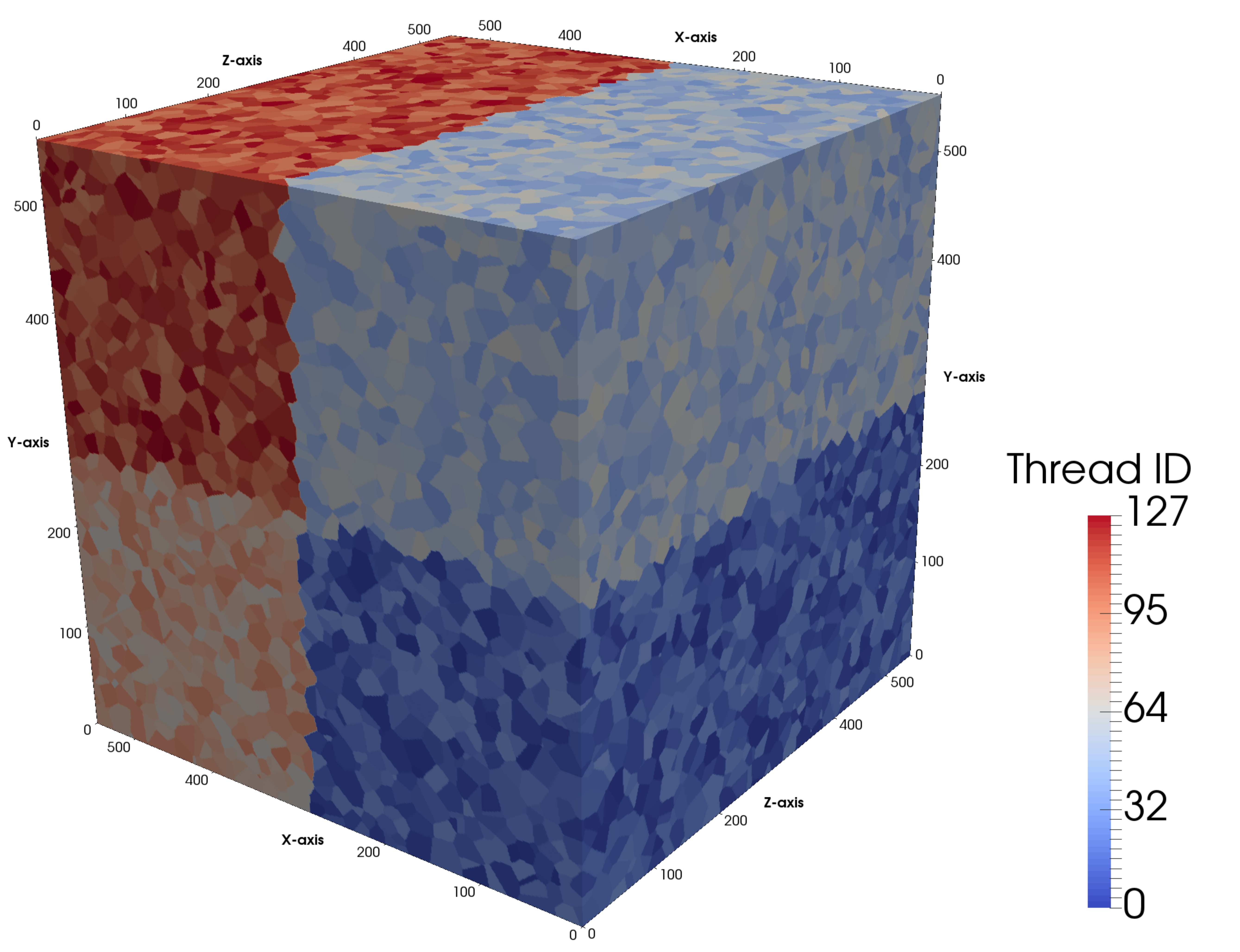} }
\caption{We illustrate the scatter of grains to computational threads by coloring grains to indicate their affiliation to a certain thread ID. We implemented two complementary concepts, the first (\cref{iterative}) aiming just for a homogeneous workload distribution and the second (\cref{squared}) also for the reduction of remote memory access during the Corrector step \cref{comparison step}.}
\label{Scheduler}
\end{minipage}
\end{figure}

Due to the grain-resolved data structure, each step of \cref{predictor} could be executed independently for each grain. Remote memory was accessed only during the \emph{corrector step} as each grain needs to refer to the $\Omega_j$s of all adjacent grains to correct its own distance function. Hence, synchronization of grains was needed after the end of each sub-routine, which meant that all grains must have completed the previous step of the algorithm with regard to the current iteration. The parallelization idea was very natural, as we just distributed grains among the nodes contributing to the simulation (\cref{Scheduler}). 

Our implementation targeted a certain computer architecture, a machine operated by \textit{BULL}, consisting of up to 128 cores coupled to one big shared memory system. Therefore, 4 physical boards owning 4 processors (Intel) were connected with BULL's Coherent Switch (BCS) chips. Thus, not only the interconnection between those 4 boards imposed a NUMA (Non Uniform Memory Access) topology but also each board consisted of four NUMA nodes connected via another QPI (Quick Path Interface). This is important to note as memory fetches from remote locations will lead to different latencies on the two level NUMAness. This issue will be further discussed in \cref{taskdistribution}. 

Essential to the parallel performance of our implementation was that each thread utilized only memory local to its corresponding core. We achieved this by employing a two step process. First, each thread was bound to a specific core using Linux's NUMA API\footnote{\url{http://linux.die.net/man/3/numa}} and affinity control API\footnote{\url{http://man7.org/linux/man-pages/man2/sched_setaffinity.2.html}}. Second, instead of using the built-in heap manager of the compiler, we used \textit{jemalloc} \cite{jemalloc}. This allocator ensured that each thread got a memory pool, in which all allocations and deallocations were executed. The allocation policy combined with the thread binding and the first touch policy of the Linux kernel allowed optimal memory placement for the simulation. 
We stress that these optimizations are not specifically bound to our algorithm and could in fact be used in any simulation that would benefit from thread local memory allocation and consumption.

The parallel performance of the \emph{predictor step} (\cref{predictor} and \cref{Gaussian}) needed certain consideration. Here, the distance function of each grain $d_i$ had to be transformed to Fourier space, where the appropriate kernel was applied as described in \cref{subSec:NumericalScheme}. To accomplish this, we utilized two libraries - FFTW \cite{FFTW05} and Intel MKL\footnote{\url{https://software.intel.com/en-us/intel-mkl}}. Both libraries require a certain precomputation effort, in the form of a \textit{computational plan} that is created before the actual transformation can take place. Due to the implementation of FFTW, this precomputation could not be executed in parallel, and thus, had to be always serialized. This serialization hindered the parallelization of the \emph{predictor step} substantially. Therefore, we turned to Intel MKL, where the plan computation could be done in parallel and no serialization was required. Furthermore, we could reuse the plans whenever the size of the bounding box had not changed with time. To save allocation effort for these out-of-place transformations each thread operated its own buffer reused by all grains in the related workload.

\subsection{Task distribution}
\label{taskdistribution}

First, a random distribution of grains to threads was implemented to achieve a homogeneous workload balance on all participating computational threads (\cref{iterative}). For all sub-routines, with the exception of the \emph{corrector step}, this strategy might be optimal as computations were limited to local memory. By contrast, the \emph{corrector step} contains a lot of remote memory access as it recomposes the network by comparing grains owned by different threads. Due to the described NUMA architecture of the hardware, these remote fetches are very expensive - up to 10-times slower than local fetches. The remote access could also be graphically identified by the amount of GBs separating differently colored grains, as we illustrated the scatter by coloring grains to indicate the affiliation to a thread ID (\cref{Scheduler}). We will evaluate the effect of this naively designed patchwork structure (\cref{iterative}) compared against the spatial scheduling (\cref{squared}) in the next chapter.

\section{Simulation setups}

\subsection{Model validation}

To validate the simulation program, a simple case of study was constructed that provides an analytical solution for the relevant driving forces. For a spherical grain embedded in a continuum, the theoretical course of the radius evolution was compared to the results of the simulation. Starting from a reasonable discretization of the spherical grain, this setup was utilized to evaluate the differences of the two competing re-initialization schemes introduced in \cref{subsec:Re-initialization}, the most advanced by Hartmann et al.\cite{Hartmann} and the fastest suggested by Zhao \cite{zhao04} to recompute the SDF to the new position of the interface after each time step. 

In addition to the curvature (also referred to as capillary) driving force, we considered a bulk energy originating from a magnetic field \cite{Molodov,Molodov20044377, Molodov200771,Molodov20103568,BarralesMora2010}.
During heat treatment of certain metallic materials with anisotropic magnetic susceptibility, such as bismuth, titanium or zirconium, the microstructure evolution can be affected by an external magnetic field as shown in several experiments \cite{Molodov20044377, Molodov200771,Molodov20103568}. The magnetic field can affect grain boundary migration \cite{Mull56} as an additional driving pressure on the GBs which emerges due to anisotropic magnetic susceptibilities causing the magnetic energy density to become dependent on the grain orientations with respect to the field direction:

\begin{equation}
\label{magneticE}
E_m \,= \, \frac{1}{2}\, \mu_0 \ H^2 \, \chi, 
\end{equation}
   
where $\mu_0,\,\chi, H$ denotes the magnetic permeability of the vacuum, the magnetic susceptibility and the magnetic field strength, respectively. According to elementary crystallography, the magnetic susceptibility of a single crystal can be written as $\chi := \left(\chi_{\perp} + \Delta \chi cos^2 \theta \right)$, where $\Delta \chi$ is the susceptibility difference of grains with their c-axis parallel $\chi_{\parallel}$ and perpendicular $\chi_{\perp}$ to the magnetic field direction. Hence, the magnetic energy density of a grain reaches its maximum when the c-axis of the grain becomes parallel to the magnetic field direction whereas it becomes minimal when its c-axis is perpendicular to it. The additional driving pressure results from the difference in the volumetric energy between two adjacent grains i, j:

\begin{equation}
\label{magneticF}
p_m \,= \, \frac{1}{2}\, \mu_0 \Delta \chi H^2 (cos^2 \theta_i  - cos^2 \theta_j).
\end{equation} 

Then, the simulation setup consisted of a spherical grain with a radius of $50\, \mu m$ and an orientation of $\left(\phi_1 = 180^{\circ}, \Phi=35^{\circ}, \phi_2=0^{\circ}\right)$ in Bunge-Euler notation. This grain was embedded in a continuum with an orientation $\left(\phi_1 = 0^{\circ}, \Phi=35^{\circ}, \phi_2=0^{\circ}\right)$. In terms of the magnetic susceptibility we took the value for titanium form literature \cite{Molodov20044377} $\Delta \chi = 1.18 \times 10^{-5}$. The magnetic field direction was chosen to be $\left(0^{\circ}, -32,8^{\circ}, 46.93^{\circ}\right)$. Thus the c-axis of the continuum or matrix grain was exactly perpendicular the the field, whereas the spherical grain's c-axis was approximately parallel to maximize the magnetic driving pressure. For various magnetic field strengths the paths of the spherical grain was tracked and compared against the corresponding analytic solution.  

\subsection{Performance analysis}

For the performance analysis of our implementation, a benchmark problem was defined, which was qualified to evaluate the performance in sequential as well as in parallel execution. For these studies, we chose the re-initialization scheme which was found to be the most beneficial. A network composed of 100,000 grains in 2D and 10,000 grains in 3D defined a rather realistic problem size which could still be computed without parallelization while the workload was sufficient to study any performance issues in parallel. The grains were initially discretized by 20 points on average per spatial direction and the networks were evolved for 20 integration time steps. 

To investigate the memory consumption and the domain superimposition, we increased the problem size in 3D to 100,000 grains and simulated the coarsening until only 100 grains remained in the system. Such a polycrystal consumed nearly 300 GB peak memory in the first time steps. It is thus sufficient in size to trace a long term evolution of the heap memory.

\subsection{Grain growth with consideration of bulk energy}

As an application of our grain growth model, the case of growth affected by a magnetic field was studied. There has been a considerable amount of work in this field providing numerous experimental results for comparison \cite{Molodov,Molodov20044377, Molodov200771,Molodov20103568}. Simulations were also performed utilizing a vertex model \cite{Barr10,BarralesMora2007160}. These simulations were capable of reproducing the essential experimental findings. It is stressed, however, that these simulations were performed only in 2D owing to the numerical complexity of implementing a vertex model in 3D \cite{L.A.BarralesMora2009,BarralesMora2010}.

A polycrystal with 500,000 grains in 3D was initialized utilizing a total number of $959^3$ grid points which equals a memory peak of about 1.2 TB. The grain orientations were randomly sampled from a reference set of orientations measured in titanium using EBSD \cite{Molodov}. The resulting texture was bimodal and very similar to the experimental one (\cref{T0Pol}). Hence, all grains had orientations close to $O1 =(180^\circ,35^\circ,0^\circ)$ or $O2=(0^\circ,35^\circ,0^\circ)$. The rolling direction (RD) of the sample was set perpendicular to the magnetic field whereas the transverse direction (TD) was tilted $32^\circ$ with respect to the field, which corresponded to a magnetic vector ($0^\circ,\textnormal{cos}(32^\circ),\textnormal{sin}(32^\circ)$). Thus, the grains either had a c-axis approximately perpendicular or parallel to the applied magnetic field. The magnetic field generated a magnetic flux density of 17 Tesla, which corresponded to the experiment carried out in \cite{Molodov}. The grains initially had a size of $d=20\,\mu m$ in diameter. Depending on crystal orientation, the grains carried an additional magnetic energy in the range of $E_m \in \left[0,1351.23\right] \frac{J}{m^3}$. We assumed a GB energy of $0.3\,J/m^2$ and a high angle GB mobility of $3 \cdot 10^{-11}m^4/Js$ for titanium. For low misorientations the GB energy depended on the disorientation angle $\theta$ between the adjacent grains following the Read-Shockley model \cite{Read50}, whereas for $\theta >15^\circ$ the GB was considered to be a high angle GB associated with a constant energy. In terms of the GB mobility, we distinguished generally between low and high angle GB's but defined a steep ascent between 8-10 degrees misorientation: 
\begin{equation}
\mu_{GB} \, =\, 1.0 - (0.99 \cdot exp(-5 \cdot \left(\frac{\theta}{15^{\circ}}\right)^{9}))
\label{eq:GB_mob}
\end{equation}
The simulations were performed until only 1000 grains were left in the microstructure.
\section{Results}
\label{graingrowth3D}

\subsection{Model validation results}
\begin{figure}[h]
\begin{minipage}{\textwidth}
\centering
	\subfloat[]{\label{redistComp} \includegraphics[height = 0.4\textwidth]{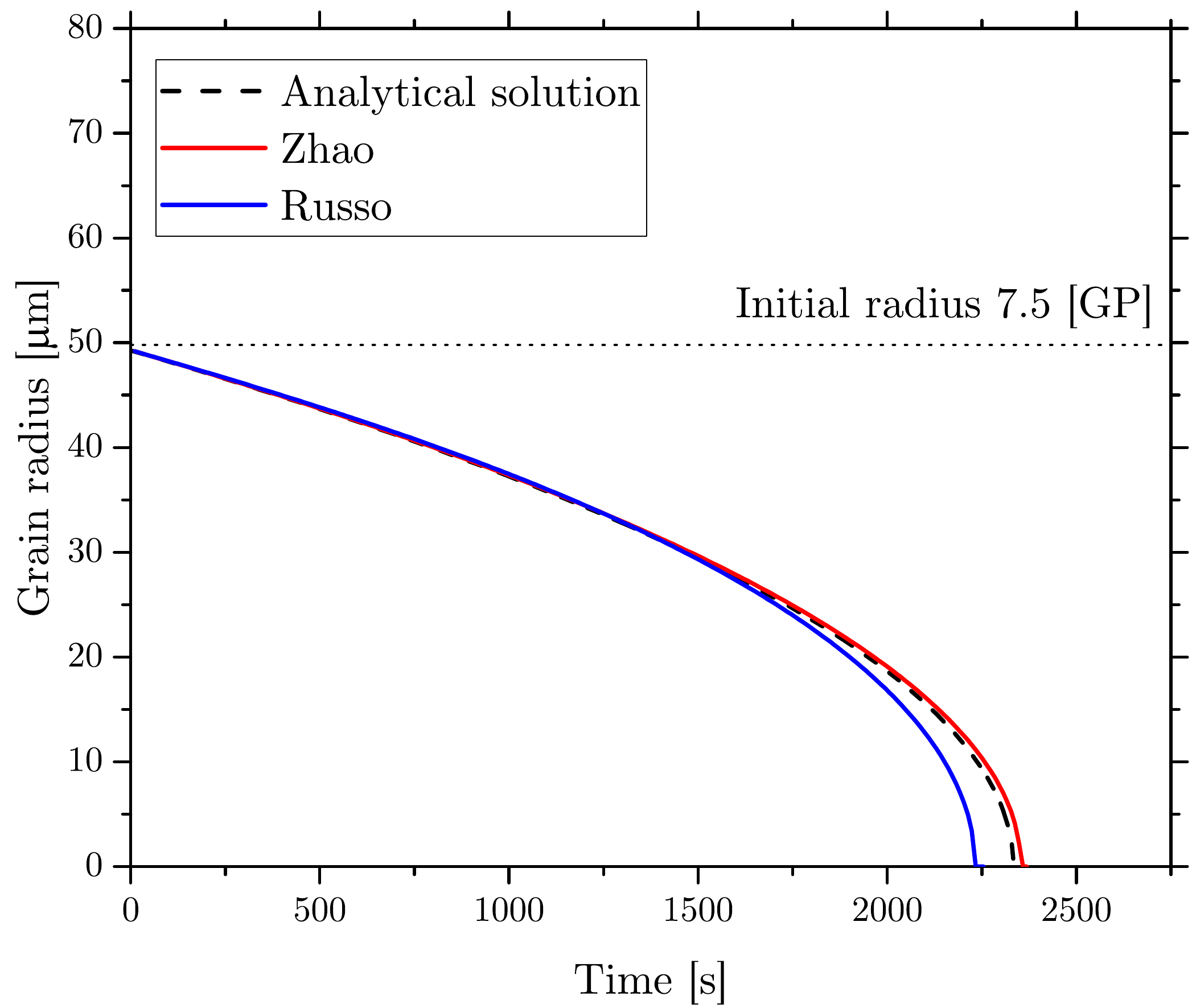} }
  \subfloat[]{\label{magneticbenchmark}  \includegraphics[height = 0.41\textwidth]{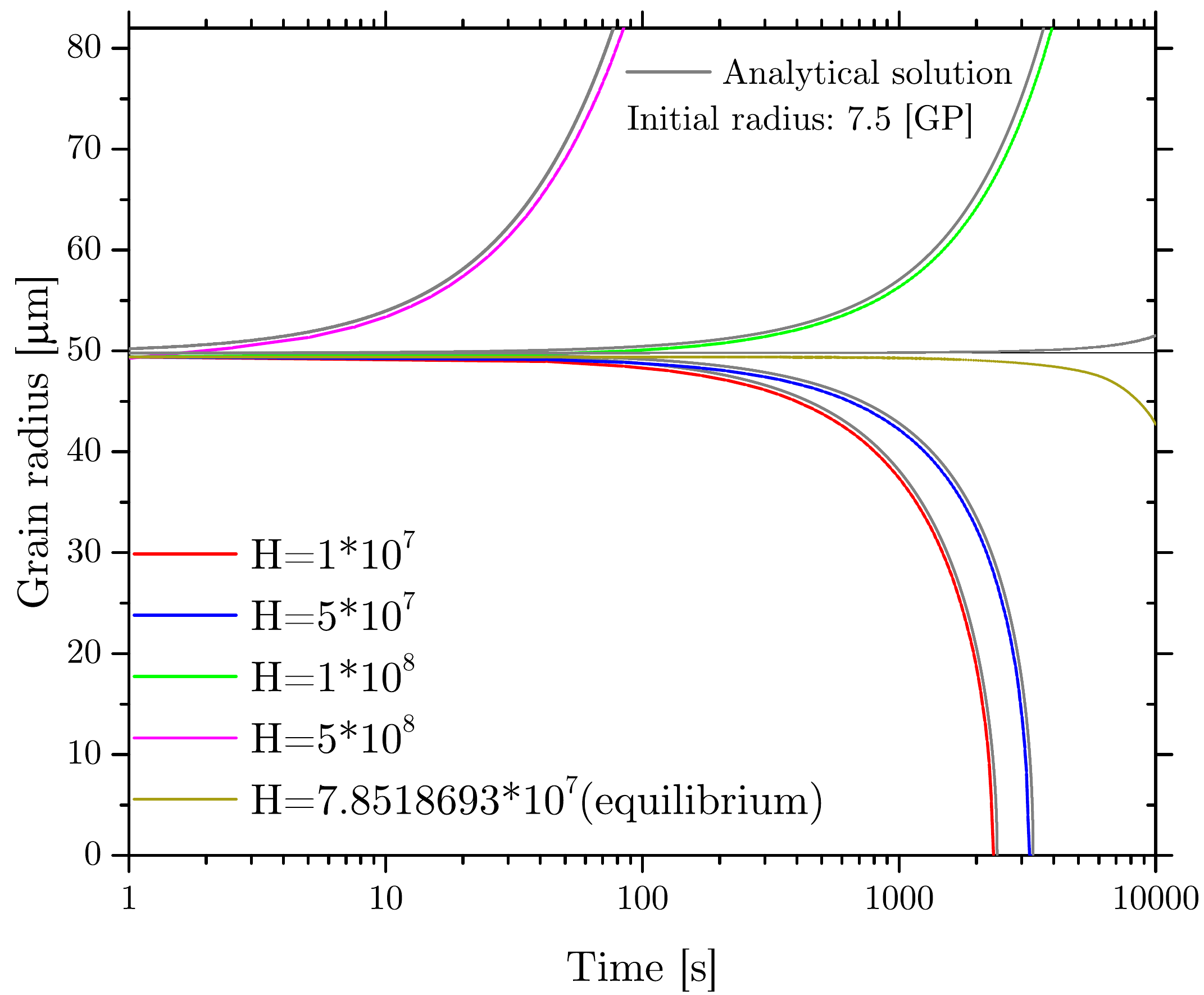} }
\caption{a) A spherical grain shrinks due to the capillary driving force utilizing two different redistancing schemes \cite{zhao04,Russ00}. b) The motion of this spherical grain, affected by various magnetic field strengths, is compared against the analytical solutions.}
\end{minipage}
\end{figure}

To begin with, we will report the findings for the spherical grain. As a mapping between physical and internal units was introduced in \cref{subsec:predictorstep}, we were able to evaluate all simulation results in physical units. In \cref{redistComp}, the shrinking grain was tracked utilizing two different re-initialization algorithms. No difference was observed until 1,500 s annealing. Reaching approximately a diameter of 7-8 grid points, an acceleration of the shrinkage was observed in the simulation, where the scheme of Hartmann et al. \cite{Hartmann} was used. In turn, the evolution of the kinetics was slightly delayed for the extremely fast method of Zhao \cite{zhao04}. Still, a better agreement in terms of the arrival time was found using the latter method.

For this reason, subsequent simulations were performed exclusively utilizing Zhao's approach. Afterward, we validated the effect of a secondary driving force originating from an external magnetic field in \cref{magneticbenchmark}. A broad range of magnetic field strengths was applied to capture even cases where the magnetic driving force exceeded the capillary pressure and forced the grain to grow. Excellent accuracy was found for all realizations of magnetic field forces.

\subsection{Performance}
\label{subsec:perf}

The program was designed to be executed on a \emph{Bull SMP-S} (BCS) machine of the RWTH-Aachen computer cluster. These machines are composed of 128 cores grouped into two NUMA levels. The processor type is an Intel Xeon X7550 with 2.00GHz clock rate (8 cores on one processor chip). Four cores on each chip share their L3 cache, whereas the L2 and L1 cache memories are individually owned by each core. Four chips (each 8 cores) are connected via a \emph{Quick Path Interface} (QPI) and form the first NUMA level. Four clusters of these $32$ core aggregates are connected with via BCS to build the entire ccNUMA architecture. The BCS connection is very similar to the QPI but slower. The RAM is limited to 128GB on the BULL SMP-S, whereas the BULL SMP-XL offers 2TB. These machines were also used to perform all of the simulations reported in this study.

\subsubsection{Sequential performance}
\label{subsec:seqperf}

\begin{figure}[h]
	\centering
	\includegraphics[width = 0.6\textwidth]{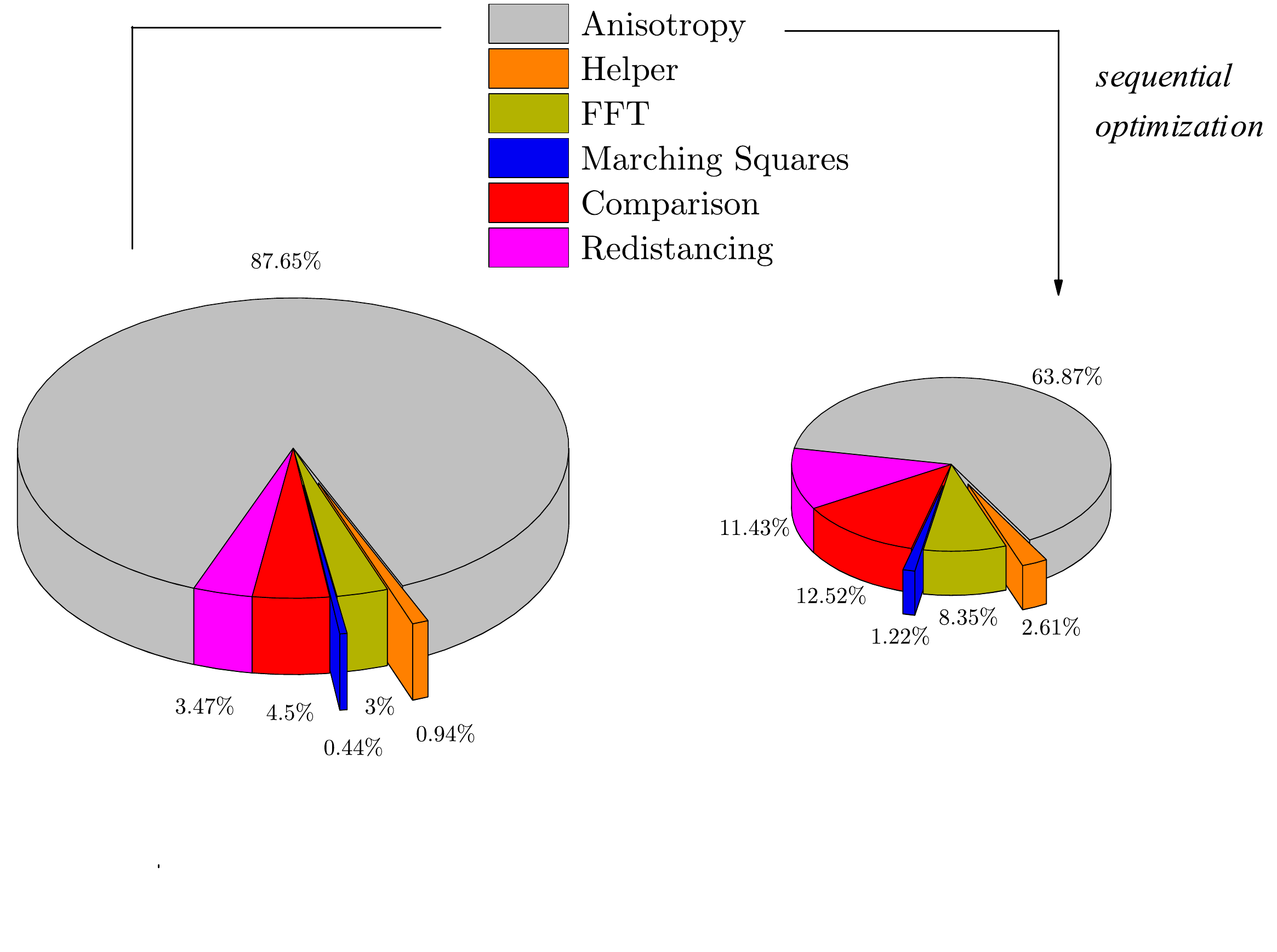}
	\caption{The shares of the subroutines during a sequential execution of the benchmark configuration computing $20$ integration steps with time for a polycrystal composed by $100,000$ grains (2D) are presented. As the convolution correction on the left turned out to be the alone-standing bottleneck of our scheme, we optimized that routine and reduced the formal complexity by a factor of four without loosing accuracy. The area of the circles correspond to the execution time of the main function (~3h vs. ~1h). The sequential run-time was improved by a factor of $2.95$ from left to right.}
\label{fig:workload}
\end{figure}

In a serial run of our benchmark example in 2D, the computational costs of each subroutine of \cref{predictor} were measured. In the initial version the anisotropy correction, introduced in detail in \cref{subsec:predictorstep,sec:seqopt}, consumed up to $87\%$ of the computational costs. As this part of the algorithm was targeted by our sequential optimizations (\cref{sec:seqopt}), the resulting benefit was evaluated in \cref{fig:workload}. By just optimizing the anisotropy correction, the run-time of the benchmark example was reduced to one third in 2D and one tenth in 3D.

\subsubsection{Parallel performance}

It is useful to clearly distinguish between savings made by the optimization of the sequential algorithm (\cref{subsec:seqperf}) and the speedup gained by parallelization. The execution of our benchmark case in 2D was repeated using 1, 16, 32, 64 and 128 cores and measured the run-time for the subroutines to obtain a detailed overview of the scaling. We will directly report the results using the two step procedure introduced in \cref{sec:parallel}, bounding threads to cores and utilizing the jemalloc allocator (as replacement of the built-in heap manager of the compiler) to assure that the first touch-policy always guarantees a local memory allocation.

\begin{figure}[h]
	\centering
	\includegraphics[width = \textwidth]{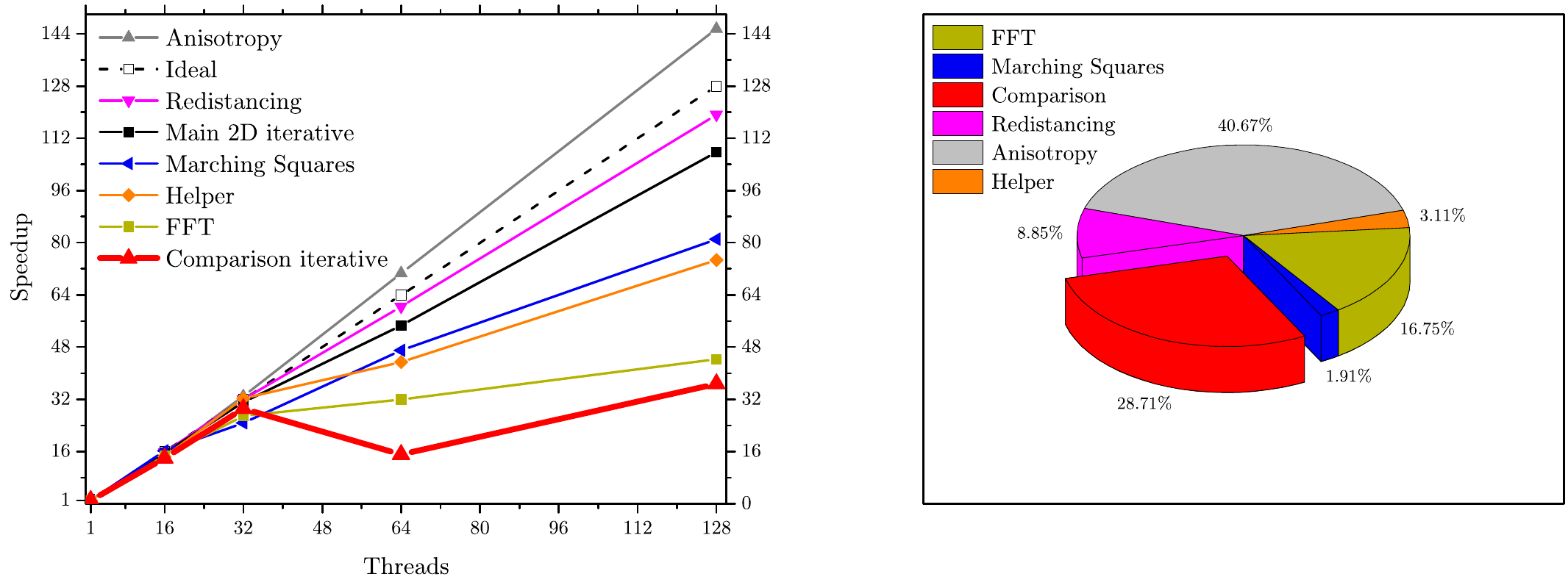}
	\caption{Strong scaling was investigated for a fixed size problem of $100.000$ grains. The main subroutines are presented individually, whereas the ideal speedup is given as reference (dashed line). While the total speedup is rather good, we achieved an acceleration of x$107.5$ utilizing 128 threads, the biggest drawback is caused by the Corrector routine. Here we observed a break down in speedup moving to the second NUMA-level of our hardware architecture.}
\label{fig:speedup}
\end{figure}

\begin{figure}[!h]
\centering
\includegraphics[width = 0.6\textwidth]{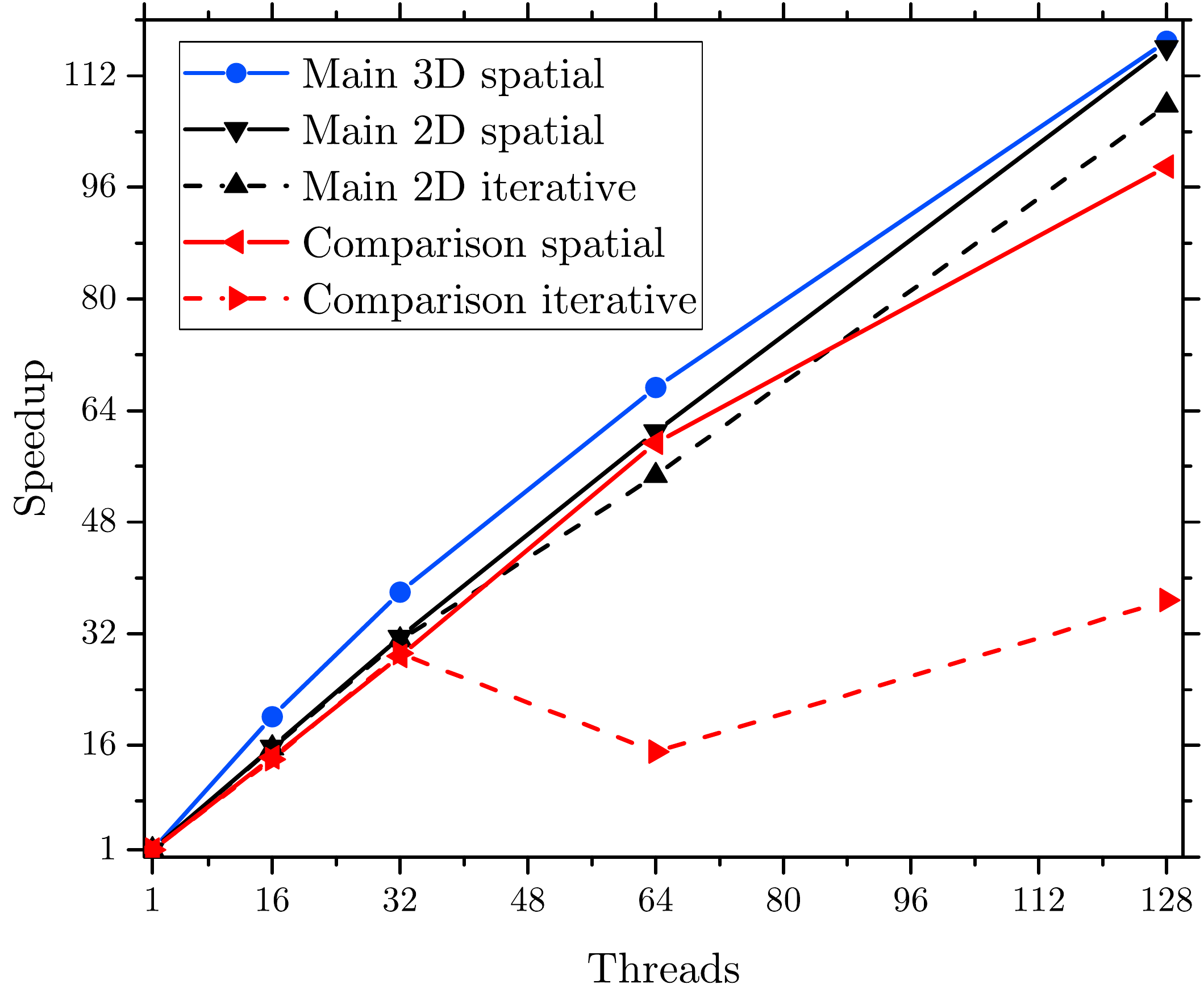}
\caption{The spatial scheduling of grains (\cref{squared}) improved the execution time of the comparison routine to average behavior. Communication in between computational threads, performed as remote memory access, did not longer delay the scaling of the algorithm. The 3D benchmark study utilizing also the spatial scheduling showed a slightly better speedup than the 2D version. Due to the parallelization the total execution times of the two benchmark cases decreased from 1h:02min to 32s in 2D respectively, from 8h:25min to 4min:19sec in 3D.}
\label{fig:superOptimized}
\end{figure}

For all subroutines, with the exception of the \emph{comparison step}, a monotonic trend in the speed-up was observed. In the best case (that of the \emph{anisotropy correction}), a super-linear acceleration of 145 times using only 128 threads was achieved, whereas the worst scaling was found for the MKL library functions resulting in a speedup of only 46. The other subroutines scaled approximately linear with different slopes. The average behavior, resulting in the total scaling and denoted in \cref{fig:superOptimized} as \textit{Main 2D iterative}, still reached a speedup of 107 compared to the sequential run time. Since the execution was limited to one NUMA node, i.e up to 32 cores \cref{fig:speedup}, linear scaling was observed also for the \emph{corrector step}. Consulting more cores, a massive delay was observed for this procedure, the only subroutine fetching memory remotely. The execution on 64 cores took even double the time compared to the one with 32 cores.

By changing the scheduling routine, we aimed to improve the scaling of the \emph{corrector procedure}. As all other subroutines access only local memory, their scaling behavior was insensitive to the scheduling procedure. For this reason, remote memory access via the BCS paths was suspected to cause the poor parallel performance. From \cref{fig:speedup}, we concluded that inter-thread communication during the comparison was not an issue as long as all threads were hosted by the same NUMA node (max. 32 threads). Therefore, we formed teams of 32 threads, hosted on the same node, to execute the workload of a spatially bounded sub-region of the entire domain (\cref{squared}). We decomposed the network into four sub-regions; equal to the number of NUMA nodes. As we explicitly pinned threads to cores, forcing the threads $\{0,\to,31\}$ to run on the first NUMA node, $\{32,\to,63\}$ on the second node and so on, we introduced a spatial scatter of threads. Aiming to reduce the remote memory access via the BCS we introduced a mapping between the physically connected regions and the teams of threads, i.e. the NUMA nodes. As the utilized allocation policy guaranteed thread-local memory allocation, this memory placement strategy hosted those grains close together in the memory which were also physically nearby (\cref{squared}).
\Cref{fig:superOptimized} illustrates the related improvements made in the critical part of the algorithm. As a result, the \emph{corrector procedure} was substantially improved reaching almost linear scaling. For the 3D benchmark case, we observed a very similar trend. The savings made by the scheduling were even more important as the data exchange rate increased exponentially owing to the higher dimensionality. 

For the \emph{anisotropy correction}, super-linear scaling behavior moved its computational cost closer to the mean (\cref{fig:speedup}). Super-linear speedup resulted from the cache effect originating from different memory hierarchies of the BCS SMP-S machine because not only the numbers of processors changed but also the size of accumulated caches from different processors and the number of memory access units. Our benchmark simulation in 2D consumed about 13 GB RAM memory, which exceeded the 8 GB memory owned by the operating processor. For this reason, the remaining memory was allocated non-local to the processor, which caused higher latencies. Hence, distributing the memory in the heap reduced the access time dramatically, which caused the extra speedup in addition to that from the actual computation. 


The total execution time for the ultimate solution in 2D was $32$ seconds after computing $20$ integration steps of the benchmark sample containing 100.000 grains and 270 s for the 3D case containing 10.000 grains.

\subsubsection{Memory consumption}
For an analysis of the memory usage, a bigger network was initialized - a 3D polycrystal composed of 100,000 grains. The simulation was run in parallel utilizing 128 threads on a BSC SMP-XL node. For all computations with memory requirements, expected to be larger than 128 GB, these jobs were scheduled to an identically constructed but taller system, a BULL SMP-XL node, with a capacity of 2TB RAM. However, the evolution of the memory consumption did not depend on the number of threads used. The results had to be classified as a sequential optimization to the model but it was hardly possible to run a production job of suitable sample size on a single core.

\begin{figure}[h]
	\centering
	\includegraphics[width = 0.6\textwidth]{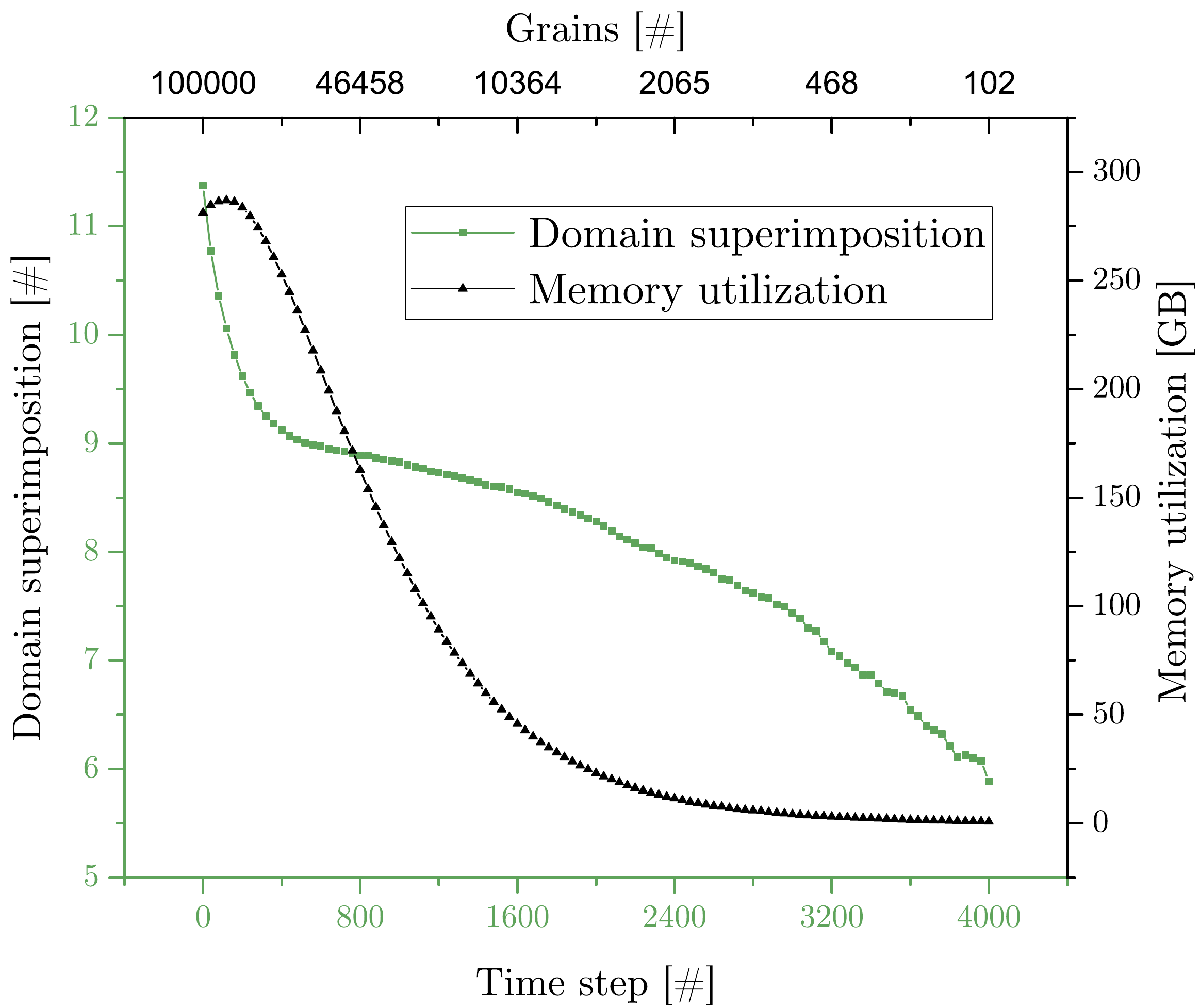}
	\caption{This plot shows the utilization of memory resources with time starting from 100,000 grains in 3D. We added up the buffer sizes in GBs and computed the superimposition of the original computational domain. The frequency of superimposition ($12 \to 6$) can be compared to former approaches, e.g. \cite{Else11} found the number to be $64$. Thus the local level-set approach saves memory by a factor of $x>6$ in 3D. Additionally, the memory utilization with time decreases nearly exponentially, as the approach consecutively benefits from the elimination of grains.}
\label{3D_memory}
\end{figure}

The memory consumption and the domain superimposition with time (\cref{3D_memory}) were studied in order to compare our implementation to global level-set (GLS) approaches from literature \cite{Bernacki15,Else13}. To determine the frequency of domain superimposition comparable to the number of GLS, we added up the number of grid points used in all our sub-domains and divided it by the size of the entire domain. \Cref{3D_memory} documents the substantial saving in memory consumption in 3D. The memory peak of the simulation was found at around 280 GB RAM. 
 
The memory consumption first increased and then dropped somehow exponentially. Surprisingly, the domain superimposition did not follow exactly the same trend but was observed to be monotonically decreasing from the beginning. We found that the volume covered by all sub-domains did not exceed 12 times the size of the entire domain in 3D.

\subsection{Large scale simulation - grain growth affected by an external magnetic field}

\begin{figure}[h]
\begin{minipage}{\textwidth}
\captionsetup[subfloat]{captionskip=1pt} 
\begin{center}
	\subfloat[]{\label{T0} \includegraphics[width = 0.33\textwidth]{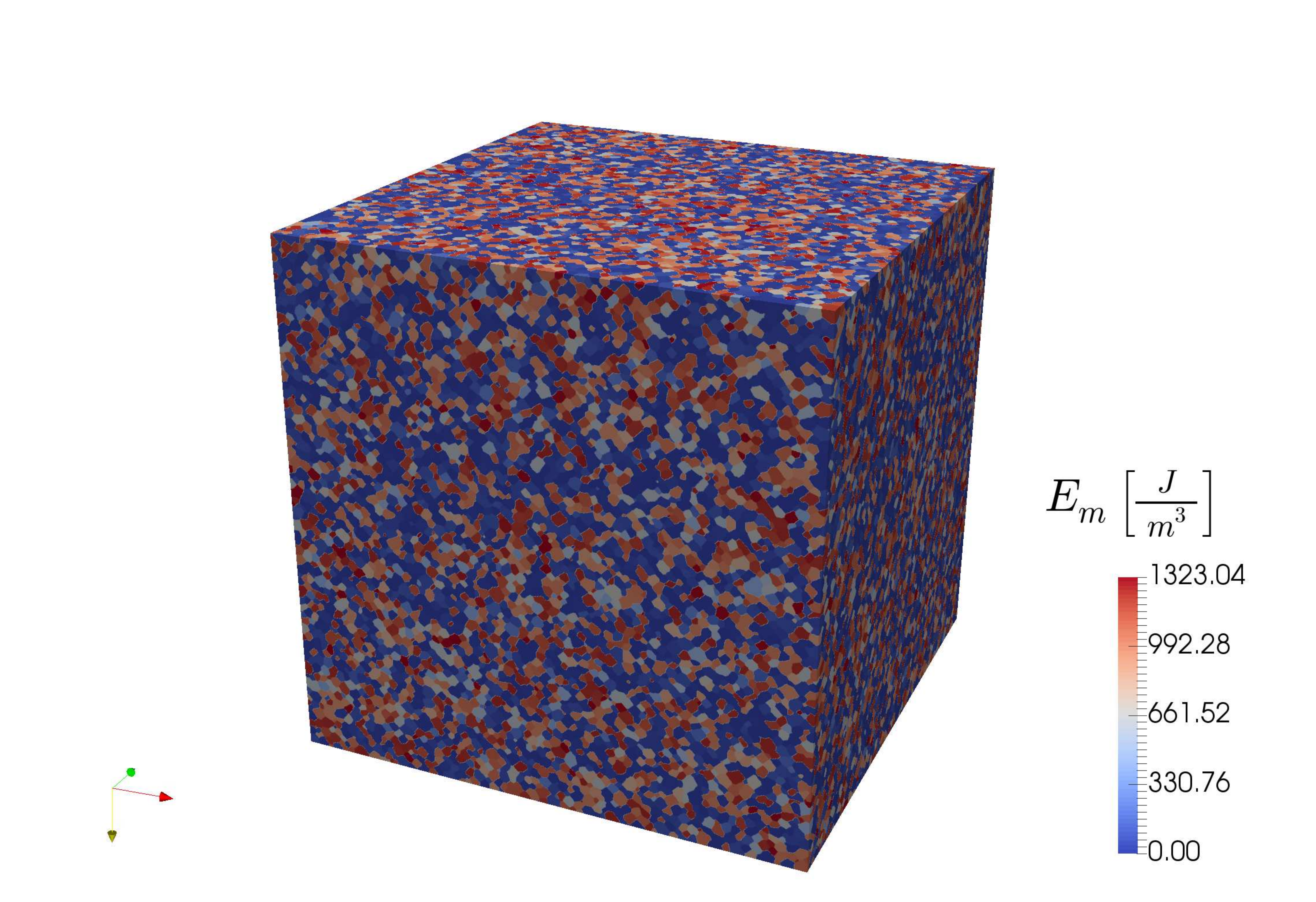} }
	\subfloat[]{\label{T1} \includegraphics[width = 0.33\textwidth]{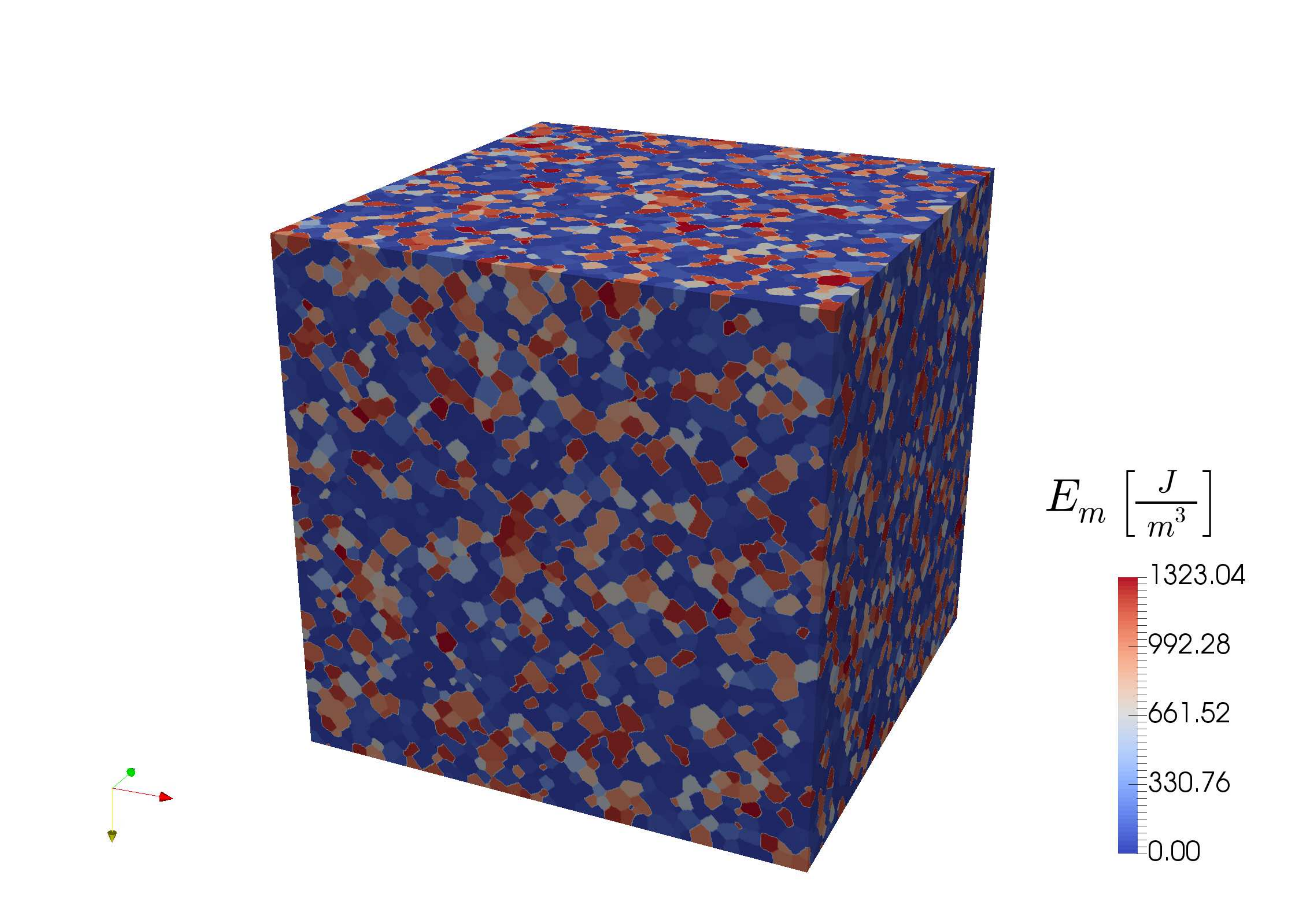} }
	\subfloat[]{\label{T2} \includegraphics[width = 0.33\textwidth]{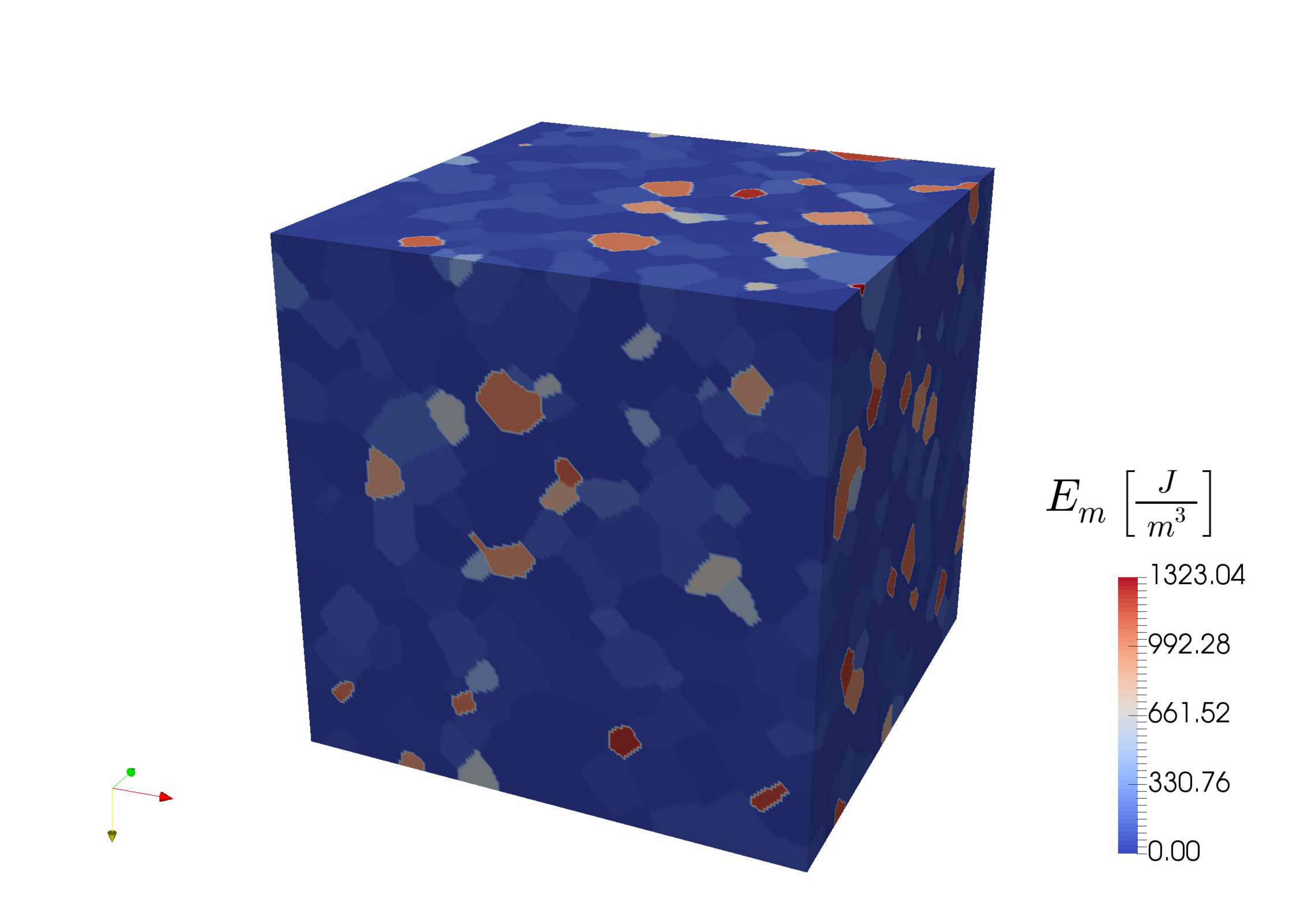} } \\
  
 	\subfloat[]{\label{T0Pol} \includegraphics[height = 0.27\textwidth]{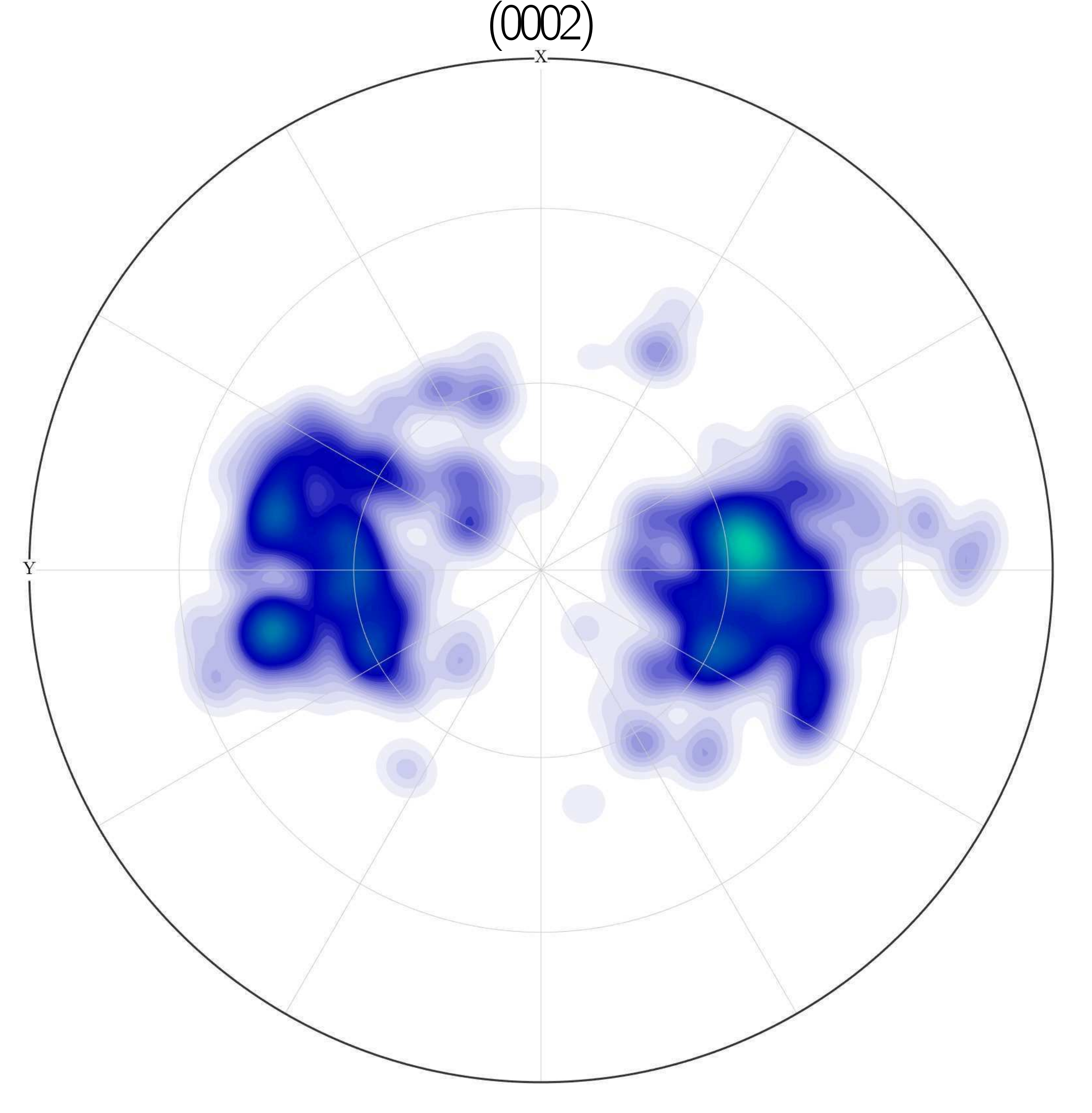} }
  \hspace{0.7cm}
	\subfloat[]{\label{T1Pol} \includegraphics[height = 0.27\textwidth]{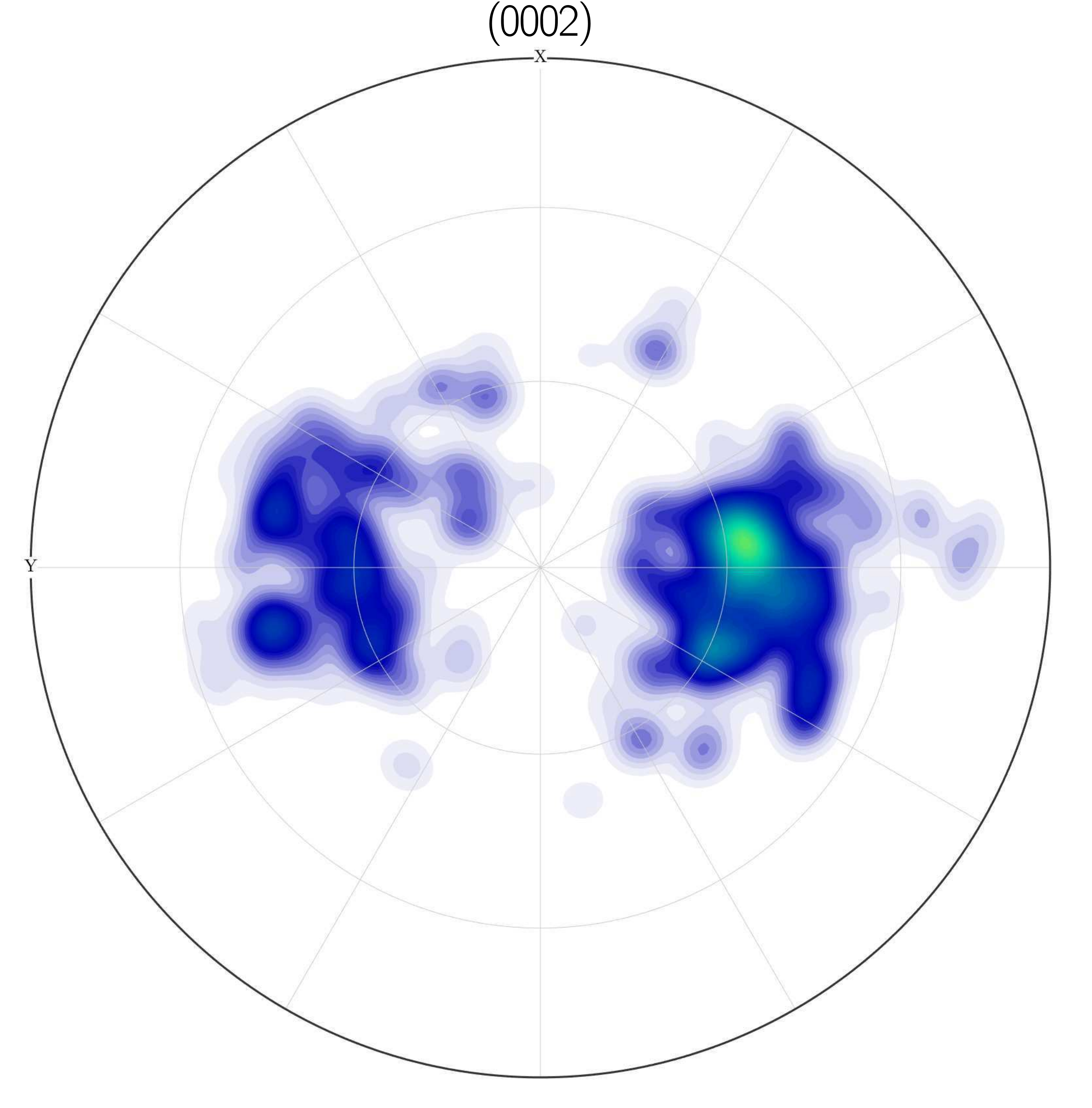} } 
  \hspace{0.7cm}
	\subfloat[]{\label{T2Pol} \includegraphics[height = 0.27\textwidth]{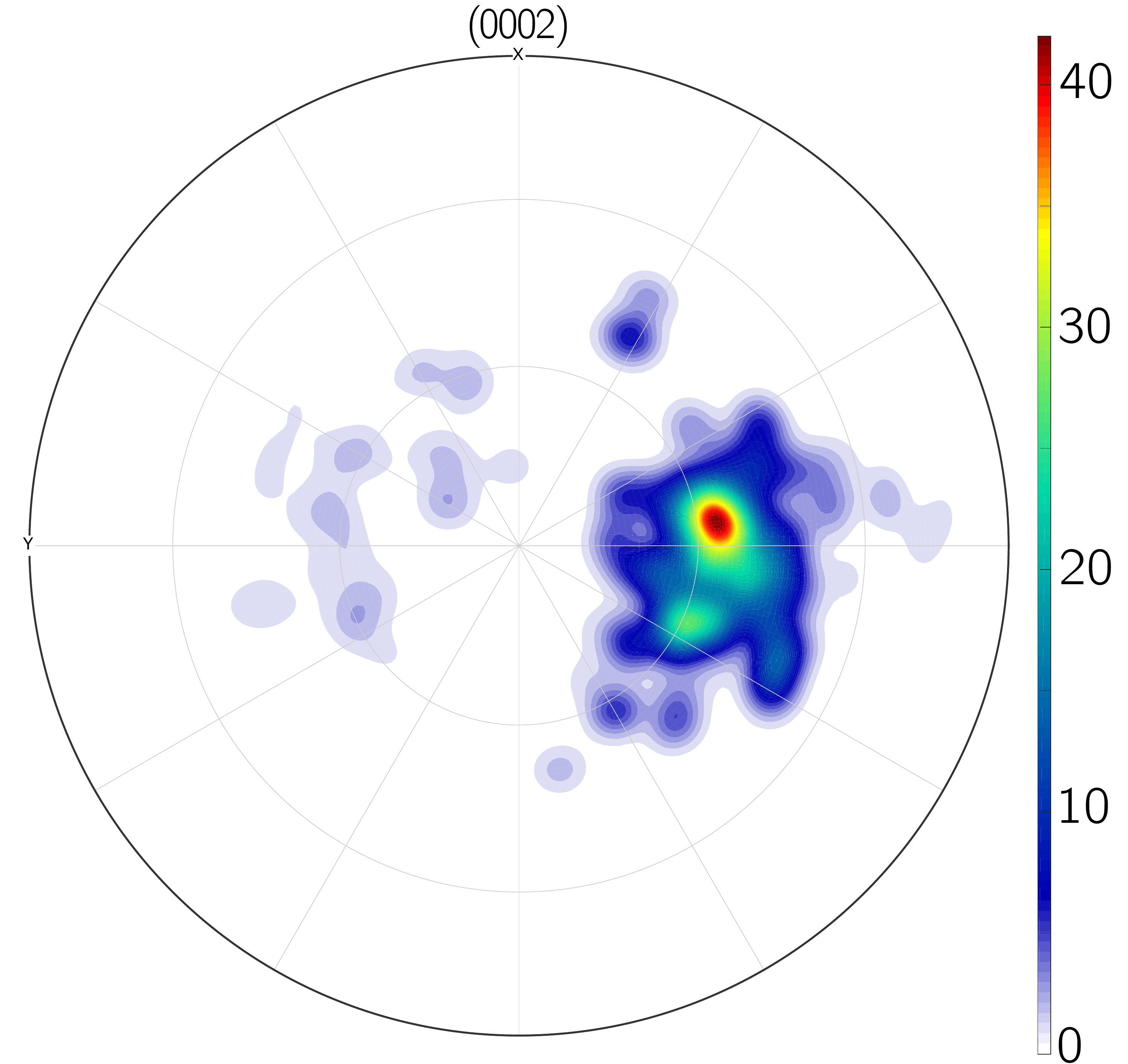} }
  \end{center}
\caption{a-c) A time series (at 10s, 55s, 500s annealing) of a polycrystal with initially $500,000$ grains on a grid of $959^3$ points was annealed under the influence of an external magnetic field until only 1,000 grains remained in the sample. Grains were colored by their magnetic energy $E_m \,[J/m^3]$. The microstructure gradually eliminated grains with a high magnetic susceptibility, colored in red, to reduce the amount of free energy in the system. d-f) The corresponding \{0002\} pole figures were plotted additionally. During annealing the disfavored texture component disappeared nearly completely.}
\label{timeseries}
\end{minipage}
\end{figure}

\begin{figure}[h]
\begin{minipage}{\textwidth}
\captionsetup[subfloat]{captionskip=1pt} 
\centering
 \subfloat[]{\label{textureKinetics}\includegraphics[height = 0.35\textwidth]{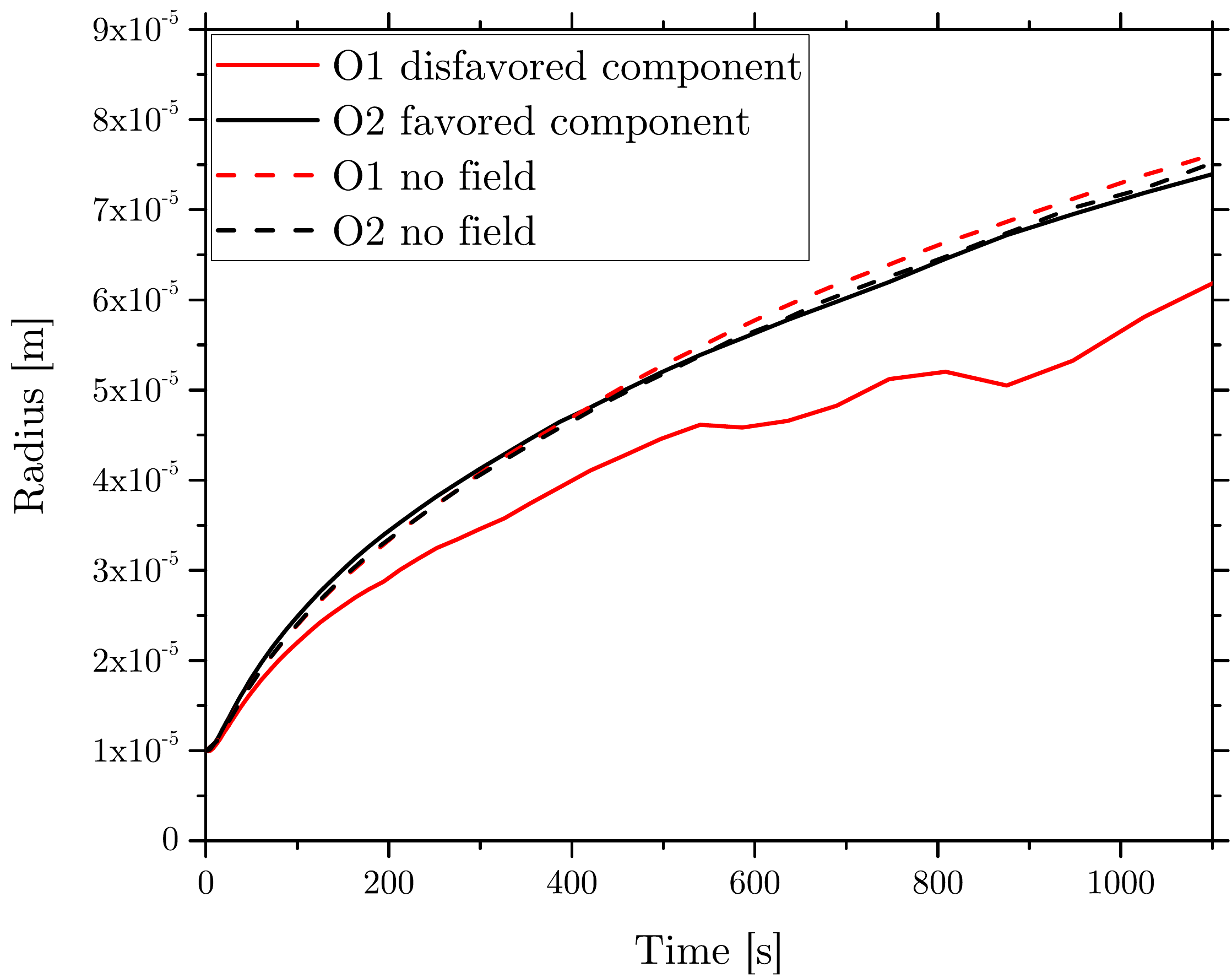} } \quad
\subfloat[]{\label{textureshares}\includegraphics[height = 0.35\textwidth]{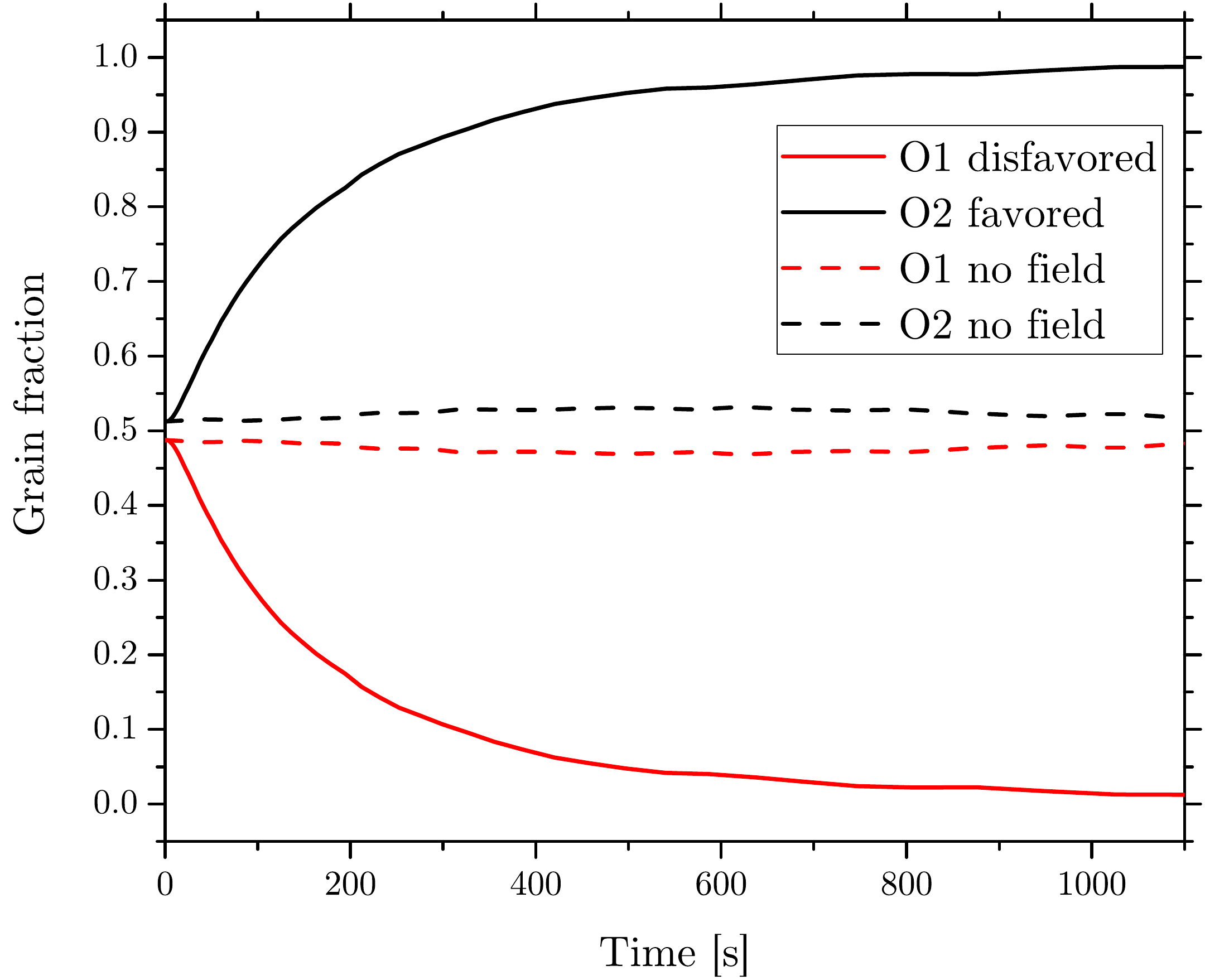} }
\caption{a) The texture evolution of the polycrystal is plotted as an evolution of the mean grain size of the preference orientations O1 and O2. Simulation of the same polycrystal without a field, indicated a strong growth selection for grains bigger than $20\mu m$ in radius. b) The fraction of favored and disfavored grains was plotted over time during magnetic annealing. }
\label{bigkinetics}
\end{minipage}
\end{figure}

We modeled a microstructure comparable to the one observed in experiments in titanium polycrystals to investigate the microstructure and texture evolution during conventional and magnetic annealing of this material (\cref{T0Pol,T1Pol,T2Pol}). Grains were clustered by their affiliation to the preferential texture components observed in experiments for rolled titanium \cite{Molodov}. Hence, we could differentiate sub-populations of grains and track their evolution with time. The evolution of the mean radius of the texture components O1 and O2, respectively which were disfavored or favored because of their different energy density according to \cref{magneticE}, was plotted in \cref{bigkinetics}. 

For comparison, the same polycrystal was simulated without the influence of a magnetic field and its mean radius evolution was added to \cref{textureKinetics}. Starting from a huge population of grains, we were able to study a long annealing time of about 1,100 s. Due to the magnetic field the volume of grains near orientation O1, whose c-axis was approximately parallel to the field, decreased continuously (\cref{T0Pol,T1Pol,T2Pol}) in presence of the magnetic field (17 T). Grains of this disfavored orientation were continuously eliminated as obvious from \cref{textureshares}. Caused by this growth selection, the final population was dominated by grains with a c-axis orientation nearly perpendicular to the applied field (\cref{T2Pol}, \cref{textureshares}). By contrast, the reference simulation (same initial polycrystal without the presence of a magnetic field) showed no texture modification in favor of a certain component (\cref{textureKinetics}). Simultaneous to the elimination of disfavored grains, the kinetics of the process slowed down. As the sub-population of favored grains increased and started occupying the whole sample volume, their size gain slowed down compared to the one observed in the reference simulation (\cref{textureKinetics}).

\section{Discussion}

\subsection{Numerical accuracy}
\label{sec:numacry}

Regarding the numerical accuracy, the effect of two competing re-initialization schemes, namely, those by Hartmann et al. \cite{Hartmann} and by Zhao \cite{zhao04} (\cref{subsec:Re-initialization}) were evaluated. Deviations from the analytical expectation for the shrinking spherical (\cref{redistComp}) grain were caused by the accumulated numerical error of the entire scheme and not just the re-initialization. The convolution itself had a numerical error as it evaluated the curvature of the isosurface depending on the resolution of the grain \cite{Esed10}. The convolution of an SDF with a Gaussian kernel acts as a smoothing of the entire SDF function. If the number of grid points separating two facing GBs becomes sufficiently small, these GBs will start attracting each other and thus, the shrinkage of the grain will be accelerated. This expected behavior was observed for the scheme of Hartmann et al. \cite{Hartmann}.

It was very surprising that the simpler re-initialization scheme by Zhao resulted in a the better agreement concerning the arrival time of the shrinking grain. The reason for this is the annihilation of numerical errors during the evaluation of the curvature and the re-initialization of the SDF. As the taxicab distance (the distance estimation in the scheme by Zhao is very similar) is always larger than the Euclidean distance, the gradient of the SDF was set to be slightly steeper than the exact solution. The subsequent convolution of a slightly distorted SDF function was certainly affected. Here, the steeper gradient of the SDF opposed the attraction of facing GBs as the resolution of the interior of the grain became very low.

Since this phenomenon is characteristic for vanishing grains, we decided to proceed with the latter scheme proposed by Zhao \cite{zhao04}. However, we want to emphasize that the arrival time is decisive for the overall kinetics of grain growth as the released volume of a shrinking grain will be occupied subsequently by the adjacent grains. Thus, the entire network kinetics would be strongly affected by this circumstance. 
Note that we do not claim that the accuracy of the Zhao's scheme is better than the one obtained with the Hartmann et al. method but it is more favorable to proceed with the Zhao scheme for three simple reasons: i) the accuracy of the \emph{entire} algorithm is higher in the case of vanishing grains, ii) the arrival time matches best and iii) the computational effort is much lower.


\subsection{Performance improvement}

The purpose of this contribution was to increase the productivity of grain growth simulations and related phenomenon considering the capabilities of modern computer architecture. We chose for a grain-related mathematical description, as provided by the level-set method, and subsequently chose an object-related implementation utilizing local level-set functions on local grids. This implementation strategy, described in \cref{sec:numinno}, was the key factor for the performance of the resulting simulation software (\cref{fig:speedup,fig:superOptimized}). The major difference between our approach and other similar models was that our data structure allowed for grain-resolved parallelism in the sense that each grain could be processed independently as an individual entity. Other approaches \cite{Else09,Else11,Else14,Hall13, Bernacki15,Scholtes2016} stored grains in GLS-functions and subsequently processed these clusters simultaneously. We believe that our approach provides the following advantages:

\begin{itemize}
\item Small sub-grids allow for precisely controlled memory layout and placement.
\item Grain-resolved parallelism enables for a flexible workload distribution among all computing threads.
\item Operations on local level-set functions have a much higher cache efficiency compared to GLS.
\item Reduced memory consumption; the frequency of domain superimposition was limited in 3D by only a factor of twelve compared to 64 GLS functions as reported in \cite{Else11} and two versus 32 in 2D.
\item Regarding each grain as a single object allows tracking the evolution of certain grains and their topological paths individually, which gives the opportunity to study the dependencies of the different macroscopic features of grains and their effect on the local topological transitions \cite{kuehbach15}.
\end{itemize}

A grain-related workload distribution among threads was not successful at once. Without a solitary thread binding policy, threads would migrate over participating cores substantially limiting the parallel performance. The migration causes expensive re-caching of memory. Still, a thread binding policy does not handle memory placement directly. A \textit{first-touch policy} combined with the built-in heap manager of the Intel compiler is supposed to allocate memory local to the operating thread but gives no guarantee in practice. A small chunk of memory can be by chance placed everywhere in the address space resulting in poorer parallel performance. To cope with this problem we turned to \textit{jemalloc}, which clusters the heap into arenas associated to a set of threads - four by default. Each arena is operated by its own heap manager and thus an allocation local to the thread was guaranteed. This coherence of task scheduling, thread binding policy and memory allocation is critical for any shared memory application. Hence, this major result regarding parallel performance (\cref{sec:parallel}) is a portable and beneficial concept for all applications capitalizing from assured thread-local memory allocation.

\subsection{Grain Growth}

\begin{figure}[h]
\label{self-similar}
\begin{minipage}{\textwidth}
\centering
	\subfloat[]{\label{grainSizedist} \includegraphics[height = 0.4\textwidth]{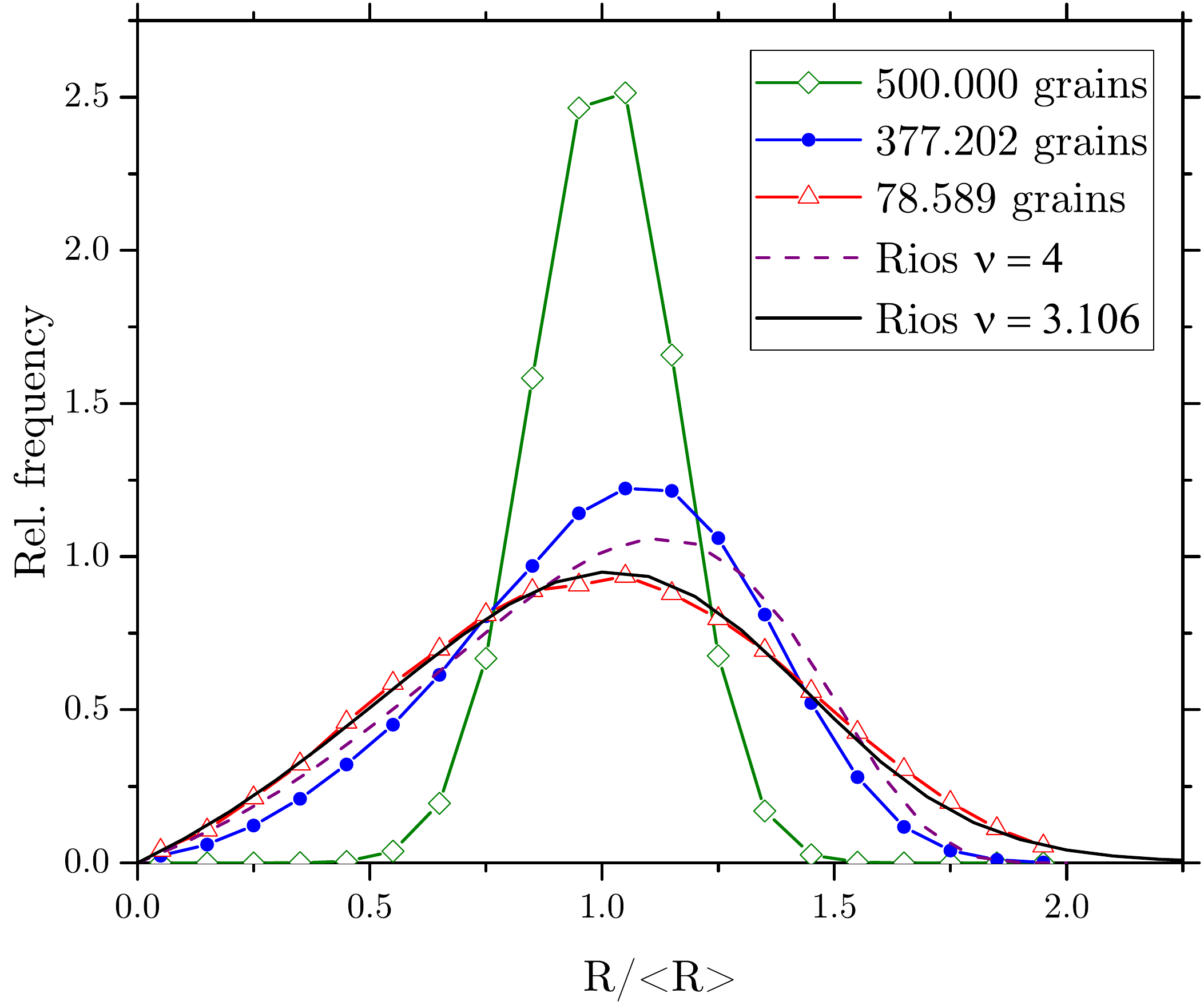} }
  \subfloat[]{\label{rhovsgrains}  \includegraphics[height = 0.4\textwidth]{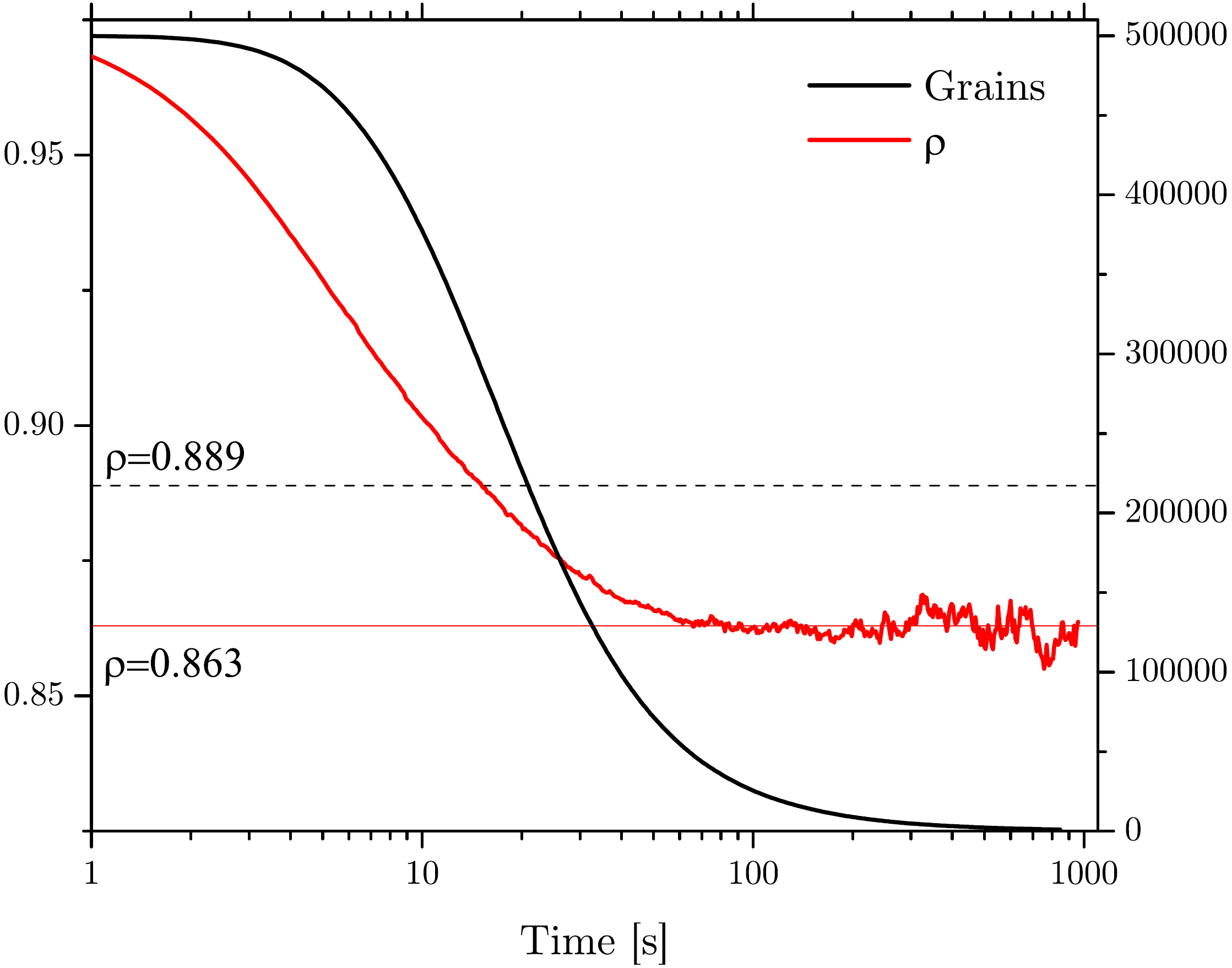} }
\caption{a) The relative grain size distribution converges to a steady state characterized by \cref{Rios} for a $\rho=0.863$. b) During the transient towards the stationary relative grain size distribution, 80\% of the initial population was eliminated. The convergence of the parameter $\rho$ indicates a steady state of the grain size distribution. A value of $\rho=0.889$, which corresponds to Hillert's GSD, does not characterize the steady state.}
\end{minipage}

\end{figure}

Regarding growth kinetics, the different LS approaches agree very well in terms of the kinetic exponent $n$ for ideal grain growth ($<R>^n-<R_0>^n = k\cdot t$). In \cite{Scholtes2016} the authors found $n=0.461$ for the whole simulation interval (initially 5,000 grains). For the steady state of a population decreasing from 500,000 grains to 1,000 grains, we observed the exponent to be $n=0.448$, whereas for the whole simulation we found $n=0.458$. The self-similar state was not reached before the initial population was thinned out to a fifth ($\sim$100,000 grains) starting also from a Voronoi-tessellation as in \cite{Scholtes2016} (\cref{grainSizedist,rhovsgrains}). The long transient is in agreement with \cite{DarvishiKamachali2015252,DarvishiKamachali20122719}, where the authors utilized a phase field approach to simulate the isotropic evolution of 30,000 grains. This very extended transient of 3D growth to approach a steady state (\cref{rhovsgrains}) demands for a huge initial population to observe a self-similar state for a suitable long period of time as fitting a growth exponent is only reasonable for the steady state.

To characterize the stationary grain size distribution a convenient formulation of the grain size distribution function (GSD) in 3D  is given by Rios et al. \cref{Rios}:

\begin{equation}
p(z,\nu) \, = \, \frac{z \rho^2 H \nu^{H0/2}}{(\rho^2 z^2 - \nu \rho z + \nu)^{1+H/2}} \cdot exp\left[ - \frac{H \nu}{\sqrt{4 \nu - \nu^2}} \left(arctan\left( \frac{2\rho z-\nu}{\sqrt{4 \nu - \nu^2}}  \right) + arctan\left( \frac{\nu}{\sqrt{4 \nu - \nu^2}} \right)    \right)    \right]
\label{Rios}
\end{equation}
\\
where $z = \frac{R}{<R>}$, $\rho = \frac{<R>}{R_{cr}}= \frac{<R>^2}{<R^2>}$= const. and $H =3$ \cite{Rios20061633,Rios20081165}. In the limit $\nu \to 4$ with $ \rho = 0.\overline{8}$, \cref{Rios} yields the Hillert distribution (\cref{grainSizedist}) \cite{HILLERT1965227}. $\rho$ characterizes the ratio between average grain size $<R>$ and the critical grain size $R_{cr}$. The latter one is mostly used for theoretical considerations and determines the variance of the distribution:

\begin{equation}
Var\left(\frac{R}{<R>}\right)\,=\,\frac{1}{<R>^2}\, \left( E(R^2) - (E(R))^2 \right) \, \approx\,\frac{1}{<R>^2}\,\left( <R^2> -<R>^2 \right)\,=\, \frac{1}{\rho} - 1
\label{varianz}
\end{equation}
\\
Furthermore, it holds for the expected value $E(z)=<\frac{R}{<R>}>=1$. The remaining parameter $\nu$ affects the skewness of the probability function \cref{Rios}. Since the variance is fully defined by $\rho$ (as $E(z)=1$), it is possible to determine the steady state from an analysis of $\rho$ alone. We stress that the Hillert distribution was neither found during the transient phase nor in steady state. For the steady state we could determine a parametrization of the distribution given in \cref{Rios} to characterize the stationary grain size distribution. Here we found $\nu = 3.106$ and $\rho =0.863$ (\cref{grainSizedist,rhovsgrains}). The results clearly indicate that for isotropic grain growth there exist a single steady state, which is perfectly characterized by the formulation proposed by Rios \cite{Rios,Rios20061633,Rios20081165}.

The greatest advantage of the level set model is that it allows simulating complex cases without the necessity of explicit handling of the topological transformations. We opted in this contribution to simulate magnetically driven grain boundary motion because the magnitude of the driving force can be calculated clearly from the physical conditions without relying on any further assumptions. In our case, the magnetic field direction was chosen in such a way that half of the population was disfavored whereas the growth of the other half was promoted. The divergence of the volume evolution of both texture components was fast for grains with a radius between $10-40\,\mu m$ until the share of O1 (approx. parallel to the field) was heavily thinned out. Afterward, grains of O2 lost their growth advantage as their neighborhood transformed and hence, these grains shared only low angle GBs with a substantially lower mobility. Finally, more than 99\% of the sample volume was occupied by grains of orientation O2. Without magnetic field, both components O1 and O2 developed in a similar fashion. The entire kinetic of the polycrystal influenced by the magnetic field dropped behind the reference one due to the loss of high angle, respectively, fast GBs. Although the magnetic field was estimated to be an order of magnitude smaller than the capillary driving force $\frac{2\cdot\gamma}{r}$, its impact was strong. However, this estimation of the capillary force is just a rough approximation and therefore it can hardly explain the growth selection in the early phase of the growth. In a polycrystal the GB curvature is much smaller than the one of an equally sized spherical shape, thus the impact of the magnetic field on the kinetic was reasonable. It is worth mentioning that the results agree very well with 2D vertex model simulations of the same phenomenon \cite{BarralesMora2007160}. This indicates in principle that the dimensionality does not affect substantially the results of the simulations as both kinetics and texture evolution were predicted by both models in excellent agreement with experiments.
\section{Conclusions}

An efficient algorithm based on the level-set approach for the simulation of 2D and 3D grain growth was presented. By optimizing the data structure and optimizing the sequential and parallel parts of the algorithm, we could speed up our simulations by three orders of magnitude. The most important innovation was the developed data structure, which divided the polycrystal into macroscopic objects - the grains. A spatial distribution of grains to computational threads nearly eliminates remote access across the shared memory space. Thus, strong scaling of our application up to 128 threads on our test architecture at the RWTH Aachen computer cluster was achieved. 
A real time scaling and rescaling of physical units was introduced allowing for easy comparability to experiments. In an application of the computational scheme, we simulated the evolution of a 3D polycrystal with 500.000 grains affected by a magnetic field. The findings on texture evolution and kinetics were compared to theoretical expectations, experimental results and previous simulations showing in all cases an excellent agreement.
\newpage

\section*{Acknowledgements} 

The authors gratefully acknowledge the financial support from the Deutsche Forschungsgemeinschaft (DFG) within the ''Reinhart Koselleck-Project'' (GO 335/44-1) as well as the FZJ\"ulich and the RWTH Aachen University for granting computing time within the frame of the JARAHPC project no 0076.

\section*{References}

\bibliography{MyEndNoteLibrary}

\begin{thebibliography}{10}
\expandafter\ifx\csname url\endcsname\relax
  \def\url#1{\texttt{#1}}\fi
\expandafter\ifx\csname urlprefix\endcsname\relax\def\urlprefix{URL }\fi
\expandafter\ifx\csname href\endcsname\relax
  \def\href#1#2{#2} \def\path#1{#1}\fi

\bibitem{hesselbarth}
H.~Hesselbarth, I.~Gobel, Simulation of recrystallization by cellular automata,
  Actae Metallurgica et Materialia 39~(9) (1991) 2135--2143.

\bibitem{brakke}
K.~A. Brakke, The surface evolver, Experimental Mathematics 1 (1992) 141--165.

\bibitem{Kuhbach2016}
M.~K\"uhbach, G.~Gottstein, L.-A. Barrales-Mora, A statistical ensemble
  cellular automaton microstructure model for primary recrystallization, Acta
  Materialia 107 (2016) 366 -- 376.

\bibitem{Kuehbach2014}
M.~K\"uhbach, L.~A. Barrales-Mora, G.~Gottstein, A massively parallel cellular
  automaton for the simulation of recrystallization, Modelling and Simulation
  in Materials Science and Engineering 22~(7) (2014) 075016.

\bibitem{chengPFM}
L.-Q. Chen, Phase-field models for microstructure evolution, Annual Review of
  Materials Research 32~(1) (2002) 113--140.

\bibitem{Stei96}
I.~Steinbach, F.~Pezzolla, B.~Nestler, M.~Seeßelberg, R.~Prieler, G.~J.
  Schmitz, J.~L.~L. Rezende, A phase field concept for multiphase systems,
  Physica D: Nonlinear Phenomena 94~(3) (1996) 135--147.

\bibitem{Kazarayanb}
A.~Kazaryan, Y.~Wang, S.~A. Dregia, B.~R. Patton, Grain growth in anisotropic
  systems: comparison of effects of energy and mobility, Acta Materialia
  50~(10) (2002) 2491--2502.

\bibitem{Kazarayn2002}
A.~Kazaryan, B.~R. Patton, S.~A. Dregia, Y.~Wang, On the theory of grain growth
  in systems with anisotropic boundary mobility, Acta Materialia 50~(3) (2002)
  499--510.

\bibitem{Else09}
M.~Elsey, S.~Esedoglu, P.~Smereka, Diffusion generated motion for grain growth
  in two and three dimensions, Journal of Computational Physics 228~(21) (2009)
  8015--8033.

\bibitem{Else11}
M.~Elsey, S.~Esedoglu, P.~Smereka, Large-scale simulation of normal grain
  growth via diffusion-generated motion, Proceedings of the Royal Society of
  London A: Mathematical, Physical and Engineering Sciences 467~(2126) (2010)
  381--401.

\bibitem{Else13}
M.~Elsey, S.~Esedoglu, P.~Smereka, Simulations of anisotropic grain growth:
  Efficient algorithms and misorientation distributions, Acta Materialia 61~(6)
  (2013) 2033--2043.

\bibitem{Else14}
M.~Elsey, S.~Esedoglu, Fast and accurate redistancing by directional
  optimization, SIAM Journal on Scientific Computing 36~(1) (2014) A219--A231.

\bibitem{Esed10}
S.~Esedoglu, S.~Ruuth, R.~Tsai, Diffusion generated motion using signed
  distance functions, Journal of Computational Physics 229~(4) (2010)
  1017--1042.

\bibitem{mypaper}
C.~Mie{\ss}en, M.~Liesenjohann, L.~A. Barrales-Mora, L.~S. Shvindlerman,
  G.~Gottstein, An advanced level set approach to grain growth – accounting
  for grain boundary anisotropy and finite triple junction mobility, Acta
  Materialia 99 (2015) 39--48.

\bibitem{Bernacki15}
B.~Scholtes, M.~Shakoor, A.~Settefrati, P.-O. Bouchard, N.~Bozzolo,
  M.~Bernacki, New finite element developments for the full field modeling of
  microstructural evolutions using the level-set method, Computational
  Materials Science 109 (2015) 388 -- 398.

\bibitem{Scholtes2016}
B.~Scholtes, R.~Boulais-Sinou, A.~Settefrati, D.~P. Muñoz, I.~Poitrault,
  A.~Montouchet, N.~Bozzolo, M.~Bernacki, 3{D} level set modeling of static
  recrystallization considering stored energy fields, Computational Materials
  Science 122 (2016) 57 -- 71.

\bibitem{Hall13}
H.~Hallberg, A modified level set approach to 2{D} modeling of dynamic
  recrystallization, Modelling and Simulation in Materials Science and
  Engineering 21~(8).

\bibitem{Kawa89}
K.~Kawasaki, T.~Nagai, K.~Nakashima, Vertex models for two-dimensional grain
  growth, Philosophical Magazine Part B 60~(3) (1989) 399--421.

\bibitem{Weyg98}
D.~Weygand, Y.~Br\'{e}chet, J.~L\'{e}pinoux, A vertex dynamics simulation of
  grain growth in two dimensions, Philosophical Magazine Part B 78~(4) (1998)
  329--352.

\bibitem{Barr07}
L.~A. Barrales-Mora, L.~S. Shvindlerman, V.~Mohles, G.~Gottstein, The effect of
  grain boundary junctions on grain microstructure evolution: 3{D} vertex
  simulation, Materials Science Forum 558-559 (2007) 1051--1056.

\bibitem{Barr08}
L.~A. Barrales-Mora, 2D and 3D grain growth modeling and simulation, Cuvillier
  Verlag, 2008.

\bibitem{Barr10}
L.~A. Barrales~Mora, 2{D} vertex modeling for the simulation of grain growth
  and related phenomena, Mathematics and Computers in Simulation 80~(7) (2010)
  1411--1427.

\bibitem{barrales1}
L.~A. Barrales~Mora, V.~Mohles, L.~S. Shvindlerman, G.~Gottstein, Effect of a
  finite quadruple junction mobility on grain microstructure evolution: Theory
  and simulation, Acta Materialia 56~(5) (2008) 1151--1164.

\bibitem{barrales2}
L.~A. Barrales-Mora, G.~Gottstein, L.~S. Shvindlerman, Effect of a finite
  boundary junction mobility on the growth rate of grains in two-dimensional
  polycrystals, Acta Materialia 60~(2) (2012) 546--555.

\bibitem{Zoel06}
D.~Z\"ollner, P.~Streitenberger, Three-dimensional normal grain growth: Monte
  carlo potts model simulation and analytical mean field theory, Scripta
  Materialia 54~(9) (2006) 1697--1702.

\bibitem{Zoel11}
D.~Z\"ollner, A potts model for junction limited grain growth, Computational
  Materials Science 50~(9) (2011) 2712--2719.

\bibitem{Zoel12}
D.~Z\"ollner, Grain microstructure evolution in two-dimensional polycrystals
  under limited junction mobility, Scripta Materialia 67~(1) (2012) 41--44.

\bibitem{SROLOVITZ19861833}
D.~Srolovitz, G.~Grest, M.~Anderson, Computer simulation of
  recrystallization—i. homogeneous nucleation and growth, Acta Metallurgica
  34~(9) (1986) 1833 -- 1845.

\bibitem{SROLOVITZ19882115}
D.~Srolovitz, G.~Grest, M.~Anderson, A.~Rollett, Computer simulation of
  recrystallization—ii. heterogeneous nucleation and growth, Acta
  Metallurgica 36~(8) (1988) 2115 -- 2128.

\bibitem{kuehbach15}
M.~K\"uhbach, L.-A. Barrales-Mora, C.~Mie{\ss}en, G.~Gottstein, Ultrafast
  analysis of individual grain behavior during grain growth by parallel
  computing, IOP Conference Series: Materials Science and Engineering 89~(1).

\bibitem{Yoshihiro07}
Y.~Suwa, Y.~Saito, H.~Onodera, Three-dimensional phase field simulation of the
  effect of anisotropy in grain-boundary mobility on growth kinetics and
  morphology of grain structure, Computational Materials Science 40~(1) (2007)
  40 -- 50.

\bibitem{Yoshihiro08}
Y.~Suwa, Y.~Saito, H.~Onodera, Parallel computer simulation of
  three-dimensional grain growth using the multi-phase-field model, Materials
  Transactions 49~(4) (2008) 704--709.

\bibitem{Nestler2005}
B.~Nestler, A 3d parallel simulator for crystal growth and solidification in
  complex alloy systems, Journal of Crystal Growth 275~(1–2) (2005) e273 --
  e278, proceedings of the 14th International Conference on Crystal Growth and
  the 12th International Conference on Vapor Growth and Epitaxy.

\bibitem{osher}
S.~Osher, J.~A. Sethian, Fronts propagating with curvature-dependent speed:
  Algorithms based on hamilton-jacobi formulations, Journal of Computational
  Physics 79~(1) (1988) 12--49.

\bibitem{Brandenburg2014294}
J.-E. Brandenburg, L.~Barrales-Mora, D.~Molodov, On migration and faceting of
  low-angle grain boundaries: Experimental and computational study, Acta
  Materialia 77 (2014) 294 -- 309.

\bibitem{Brandenburg2013980}
J.-E. Brandenburg, L.~Barrales-Mora, D.~Molodov, G.~Gottstein, Effect of
  inclination dependence of grain boundary energy on the mobility of tilt and
  non-tilt low-angle grain boundaries, Scripta Materialia 68~(12) (2013) 980 --
  983.

\bibitem{BarralesMora2016179}
L.~A. Barrales-Mora, D.~A. Molodov, Capillarity-driven shrinkage of grains with
  tilt and mixed boundaries studied by molecular dynamics, Acta Materialia 120
  (2016) 179 -- 188.

\bibitem{1757-899X-89-1-012008}
D.~A. Molodov, L.~A. Barrales-Mora, J.-E. Brandenburg, Grain boundary motion
  and grain rotation in aluminum bicrystals: recent experiments and
  simulations, IOP Conference Series: Materials Science and Engineering 89~(1)
  (2015) 012008.

\bibitem{barrales-mora2016}
L.~Barrales-Mora, D.~A. Molodov, J.~E. Brandenburg, Effect of grain boundary
  geometry on grain rotation during curvature-driven grain shrinkage, Diffusion
  Foundations 9 (2016) 73--81.

\bibitem{hoffrogge2016kinetic}
P.~W. Hoffrogge, L.~A. Barrales-Mora, Grain-resolved kinetics and rotation
  during grain growth of nanocrystalline aluminium by molecular dynamics,
  Computational Materials Science 128 (2017) 207--222.

\bibitem{ZhaoMerriOsher}
H.-K. Zhao, T.~Chan, B.~Merriman, S.~Osher, A variational level set approach to
  multiphase motion, Journal of Computational Physics 127~(1) (1996) 179 --
  195.

\bibitem{Zhao_variationalformulation}
H.-K. Zhao, S.~Osher, T.~Chan, B.~Merriman, Variational formulation for motion
  of multiple junctions and interfaces by level set approach.

\bibitem{Evans}
L.~Evans, Partial Differential Equations, American Mathematical Society, 1998.

\bibitem{Nemitz}
O.~Nemitz, Anisotrope {V}erfahren in der {B}ildverarbeitung:
  Gradientenfl\"usse, {L}evel-{S}ets und {N}arrow {B}ands, Dissertation,
  Universit\"at Bonn,
  \url{http://numod.ins.uni-bonn.de/research/papers/public/Ne08.pdf} (2008).

\bibitem{Hartmann}
D.~Hartmann, M.~Meinke, W.~Schröder, Differential equation based constrained
  reinitialization for level set methods, Journal of Computational Physics
  227~(14) (2008) 6821--6845.

\bibitem{Russ00}
G.~Russo, P.~Smereka, A remark on computing distance functions, Journal of
  Computational Physics 163~(1) (2000) 51--67.

\bibitem{Sussman}
M.~Sussman, P.~Smereka, S.~Osher, A level set approach for computing solutions
  to incompressible two-phase flow, Journal of Computational Physics 114~(1)
  (1994) 146--159.

\bibitem{zhao04}
H.~Zhao, A fast sweeping method for eikonal equations, Mathematics of
  Computation 74~(250) (2004) 603–627.

\bibitem{FFTW05}
M.~Frigo, S.~G. Johnson, The design and implementation of {FFTW3}, Proceedings
  of the IEEE 93~(2) (2005) 216--231, special issue on ``Program Generation,
  Optimization, and Platform Adaptation''.

\bibitem{intelMKL}
Intel Math Kernel Library. Reference Manual, Intel Corporation, 2009.

\bibitem{eigenweb}
G.~Guennebaud, B.~Jacob, et~al., Eigen v3, http://eigen.tuxfamily.org (2010).

\bibitem{Guttman:1984:RDI:971697.602266}
A.~Guttman, R-trees: A dynamic index structure for spatial searching, SIGMOD
  Rec. 14~(2) (1984) 47--57.

\bibitem{marchingCubes}
D.~A. Rajon, W.~E. Bolch, Marching cube algorithm: review and trilinear
  interpolation adaptation for image-based dosimetric models, Computerized
  Medical Imaging and Graphics 27~(5) (2003) 411--435.

\bibitem{jemalloc}
J.~Evans, A scalable concurrent malloc(3) implementation for freebsd (2006).

\bibitem{Molodov}
D.~A. Molodov, C.~Bollmann, P.~J. Konijnenberg, L.~A. Barrales-Mora, V.~Mohles,
  Annealing texture and microstructure evolution in titanium during grain
  growth in an external magnetic field, MATERIALS TRANSACTIONS 48~(11) (2007)
  2800--2808.

\bibitem{Molodov20044377}
D.~Molodov, A.~Sheikh-Ali, Effect of magnetic field on texture evolution in
  titanium, Acta Materialia 52~(14) (2004) 4377 -- 4383.

\bibitem{Molodov200771}
D.~Molodov, C.~Bollmann, G.~Gottstein, Impact of a magnetic field on the
  annealing behavior of cold rolled titanium, Materials Science and
  Engineering: A 467~(1–2) (2007) 71 -- 77.

\bibitem{Molodov20103568}
D.~Molodov, N.~Bozzolo, Observations on the effect of a magnetic field on the
  annealing texture and microstructure evolution in zirconium, Acta Materialia
  58~(10) (2010) 3568 -- 3581.

\bibitem{BarralesMora2010}
L.~A. Barrales-Mora, 2{D} vertex modeling for the simulation of grain growth
  and related phenomena, Mathematics and Computers in Simulation 80~(7) (2010)
  1411--1427.

\bibitem{Mull56}
W.~W. Mullins, Two‐dimensional motion of idealized grain boundaries, Journal
  of Applied Physics 27~(8) (1956) 900--904.

\bibitem{BarralesMora2007160}
L.~Barrales-Mora, V.~Mohles, P.~Konijnenberg, D.~Molodov, A novel
  implementation for the simulation of 2-{D} grain growth with consideration to
  external energetic fields, Computational Materials Science 39~(1) (2007) 160
  -- 165, {P}roceedings of the 15th International Workshop on Computational
  Mechanics of Materials.

\bibitem{L.A.BarralesMora2009}
L.~A. Barrales-Mora, V.~Mohles, G.~Gottstein, L.~S. Shvindlerman, Network and
  vertex models for grain growth, in: ASM Handbook: Fundamentals of Modeling
  for Metals Processing, Vol. 22a, ASM International, 2009, pp. 266--290.

\bibitem{Read50}
W.~T. Read, W.~Shockley, Dislocation models of crystal grain boundaries,
  Physical Review 78~(3) (1950) 275--289.

\bibitem{DarvishiKamachali2015252}
R.~D. Kamachali, A.~Abbondandolo, K.~Siburg, I.~Steinbach, Geometrical grounds
  of mean field solutions for normal grain growth, Acta Materialia 90 (2015)
  252 -- 258.

\bibitem{DarvishiKamachali20122719}
R.~D. Kamachali, I.~Steinbach, 3-d phase-field simulation of grain growth:
  Topological analysis versus mean-field approximations, Acta Materialia
  60~(6–7) (2012) 2719 -- 2728.

\bibitem{Rios20061633}
P.~Rios, T.~Dalpian, V.~Brandão, J.~Castro, A.~Oliveira, Comparison of
  analytical grain size distributions with three-dimensional computer
  simulations and experimental data, Scripta Materialia 54~(9) (2006) 1633 --
  1637.

\bibitem{Rios20081165}
P.~Rios, M.~Glicksman, Polyhedral model for self-similar grain growth, Acta
  Materialia 56~(5) (2008) 1165 -- 1171.

\bibitem{HILLERT1965227}
M.~Hillert, On the theory of normal and abnormal grain growth, Acta
  Metallurgica 13~(3) (1965) 227 -- 238.

\bibitem{Rios}
P.~R. Rios, M.~E. Glicksman, Topological theory of abnormal grain growth, Acta
  Materialia 54~(19) (2006) 5313--5321.

\end{thebibliography}

\end{document}